\documentclass[letterpaper,12pt]{article}

\usepackage{amssymb,latexsym,amsmath}
\usepackage{colortbl}%colorize tables or pieces of textx (gives problems)

\usepackage{fullpage}
\usepackage{nicefrac}

\usepackage{graphicx}

\makeatletter \@addtoreset{equation}{section}
\makeatother

\begin{document}
\begin{titlepage}
	\thispagestyle{empty}
	\begin{flushright}
		\hfill{DFPD-10/TH/4}\\
		\hfill{LPTENS 10/15}
	\end{flushright}
	
	\vspace{35pt}  
	 
	\begin{center}
	    { \LARGE{\bf Type IIB supergravity on \\[5mm]
	squashed Sasaki--Einstein manifolds}}
		
		\vspace{50pt}
		
		{Davide Cassani$^a$, Gianguido Dall'Agata$^{a,b,c}$ and Anton F. Faedo$^{b}$}
		
		\vspace{25pt}
		
		{\small
		{\it ${}^a$  Dipartimento di Fisica ``Galileo Galilei''\\
		Universit\`a di Padova, Via Marzolo 8, 35131 Padova, Italy}
		
		\vspace{15pt}
		
		{\it ${}^b$  INFN, Sezione di Padova \\
		Via Marzolo 8, 35131 Padova, Italy}
		
		\vspace{15pt}
		
		{\it ${}^c$ Laboratoire de Physique Th\'eorique de l'Ecole Normale Sup\'erieure \\
		24 rue Lhomond, 75231 Paris Cedex 05, France}\\

		\vspace{15pt}

		{\tt cassani, dallagat, faedo AT pd.infn.it}
		}
		
		\vspace{35pt}
		
		%{ABSTRACT}

\begin{abstract}	
We provide a consistent ${\cal N}=4$ Kaluza--Klein truncation of type IIB supergravity on general 5-dimensional squashed Sasaki--Einstein manifolds. 
Our reduction ansatz keeps all and only the supergravity modes dual to the universal gauge sector of the associated conformal theories, via the gauge/gravity correspondence.
The reduced \hbox{5-dimensional} model displays remarkable features: it includes both zero-modes as well as massive iterations of the Kaluza--Klein operators on the internal manifold; it contains tensor fields dual to vectors charged under a non-abelian gauge group; it has a scalar potential with a  non-supersymmetric AdS vacuum in addition to the supersymmetric one.
\end{abstract}
	
\end{center}
	
\vspace{10pt}

\end{titlepage}

\baselineskip 6 mm

%\date{}

%%%%%%%%%%%%%%%%%%%%%%%%%%%%%%%%%%%%%%%%%%%%%%%%%%%%%%%%%%%%%%
%%%%%%%%%%%%%%%%%%%%%%%%%%%%%%%%%%%%%%%%%%%%%%%%%%%%%%%%%%%%%%

%%%%%%%%%%%%%%%%%%%%%%%%%%%%%%%%%%%%%%%%%%%%%%%%%%%%%%%%%%%%%%
\section{Introduction} % (fold)
\label{sec:introduction}

Since the advent of the gauge/gravity correspondence, 5-dimensional supergravity played a prominent role in understanding strong coupling effects in 4-dimensional gauge theories.
Although many interesting physical effects could be captured by studying purely 5-dimensional models, a consistent analysis generically requires the knowledge of the underlying 10-dimensional theory, of its Kaluza--Klein (KK) spectrum and of its non-linear interactions.
Consistent truncations provide an efficient approach to take into account only a finite number of states in the effective theory and hence to consider purely 5-dimensional models, ensuring at the same time the lift of all their solutions to the higher-dimensional theory. 
It is therefore extremely interesting to provide classes of such models, possibly detailing the full non-linear reduction ansatz.

A key example in this context is ${\cal N} = 8\,$ SO(6) gauged supergravity, which can be obtained by restriction to the massless modes of type IIB supergravity compactified on the 5-sphere~$S^5$.
This model is believed to be a consistent truncation, although a complete proof of consistency of the reduction has yet to be given (see for instance \cite{Cvetic:2000nc} for a discussion on the consistency of the full ${\cal N} = 8$ and \cite{Tsikas,Lu:1999bw} for consistent embeddings of further truncations of the massless spectrum). 
However, many interesting deformations and solutions of the dual ${\cal N} =4$ super-Yang--Mills theory have been addressed in this 5-dimensional context, and it is obviously desirable to have similar results for more general and less supersymmetric models.

Type IIB anti-de Sitter (AdS) vacua preserving 1/4 of supersymmetry can be constructed by employing Sasaki--Einstein (SE) manifolds as internal spaces.
There are nowadays infinite classes of such manifolds with explicitly known metrics (the $Y^{p,q}$ and $L^{p,q,r}$ spaces, in addition to the $T^{1,1}$ coset space and the sphere $S^5$), and all of them  lead to different dual ${\cal N} = 1$ super-Yang--Mills theories coupled to matter.
Unfortunately, for these compactifications one cannot consistently perform massless truncations retaining the non-abelian gauge symmetries~\cite{Hoxha:2000jf}.
This has led to consistent truncations where only the supergravity multiplet is retained out of the massless modes \cite{BuchelLiu,Gauntlett:2007ma} and therefore most of the interesting physics related to the gauge and flavour symmetry of the dual theories is lost.

This limited setup can be overcome by including massive modes \cite{Bremer:1998zp,MaldMartTach}, possibly singlets of some symmetry of the internal manifold.
Including massive modes is indeed crucial for obtaining new interesting physics, like the construction of string theory backgrounds with non-relativistic conformal symmetry \cite{MaldMartTach} and emergent relativistic conformal symmetry in superfluids or superconducting states of strongly coupled gauge theories \cite{Gubser1, Gubser2}.

In this paper we broaden these results by providing a consistent ${\cal N} = 4$ (half-maximal) Kaluza--Klein truncation of type IIB supergravity on general 5-dimensional squashed SE manifolds.
Besides including the modes discussed in \cite{Bremer:1998zp,BuchelLiu,MaldMartTach,Gubser1}, our truncation incorporates in a supersymmetric way all the supergravity states dual to the universal gauge sector of the associated conformal theories.
Whatever the matter content and flavour symmetry of the dual field theory, one can always consider operators constructed by polynomials of the super-Yang--Mills multiplet.
Due to the fermionic nature of the corresponding superfield $W_\alpha$, only few such states can be constructed and, being $W_\alpha$ a singlet of the flavour group, the truncation to these states allows for a consistent reduction to a model with a finite number of states.

From a technical point of view we work along the line of similar reductions performed on 7-dimensional SE manifolds \cite{Gauntlett:2009zw,Gauntlett2}. 
We exploit the structure group of SE manifolds by expanding the 10-dimensional fields in terms of the differential forms defining the structure itself (hence singlets of the structure group).
This basis of forms is a closed system under exterior differentiation and Hodge duality, and contains all the necessary information to describe the metric sector.
This is all that we will need in order to prove that the resulting effective action is a consistent truncation of type IIB supergravity.
Our truncation is actually a reduction on squashed SE manifolds, because not only we keep the overall volume mode, but we also allow for a squashing mode between the U(1) fibre and the K\"ahler--Einstein base of SE manifolds. 

Although the above discussion may lead to the expectation that our truncation should provide minimal supersymmetric theories in 5 dimensions (as also argued in \cite{Gauntlett:2009zw}), the reduction process actually retains another gravitino, which is part of a massive multiplet at the supersymmetric SE vacuum.
We will in fact show that our 5-dimensional model fits nicely into the general description of ${\cal N} = 4$ gauged supergravity theories \cite{DHZ,SchonWeidner}, with the gauging process yielding partial or complete spontaneous supersymmetry breaking.
From the $\mathcal N=4$ point of view, our model includes 2 vector multiplets in addition to the gravity multiplet.
It encompasses previous truncations to the breathing mode, to pure ${\cal N} = 2$ supergravity and to the non-supersymmetric massive truncation of \cite{MaldMartTach}.
Still, this model cannot be obtained from any truncation of the ${\cal N} =8$ theory.

The 10-dimensional non-trivial geometry and the 5-form flux induce a gauging on the effective 5-dimensional theory that can be described by a gauge group $G$ being the product of the simplest Heisenberg group with a U(1) $R$-symmetry: $G =$ Heis$_3 \times$U(1)$_R$.
All the 8 vector fields of the ungauged theory are in a non-trivial representation of the gauge group and this implies that 4 of them need to be dualized to tensor fields.
This is a remarkable realization of a possibility first analyzed in the context of 5-dimensional models in \cite{Bergshoeff:2002qk}, i.e.~tensor fields coming from the dualization of vector fields in a non-trivial representation of a non-abelian gauge group.
Gauging an extended supergravity requires a non-trivial scalar potential, which is also present in our reduction.
The analysis of this potential shows that our truncations always admit two distinct AdS critical points, a supersymmetric (round) one and a non-supersymmetric (squashed) one.
We computed the masses of the scalar fields, as well as those of the vector fields, at these vacua and found that there are no unstable modes surviving our truncation.
We must point out, however, that the non-supersymmetric critical point can be related to the Pope--Warner deformation \cite{RomansIIBsols,Pope:1984bd} of SE manifolds. 
Hence tachyonic modes may arise along directions that we have truncated out, as it is known to be the case when the internal SE manifold is $S^5$ \cite{Zaffa}.

As follows from this presentation, our action offers a number of possible applications in the context of the AdS/CFT correspondence, incorporating previous results \cite{MaldMartTach,Gubser1,Gubser2} and enlarging the spectrum of possible 5-dimensional models and solutions that can be exactly embedded in type IIB supergravity. 

The paper is organized in four parts. 
We describe the SE structure and the necessary techniques to perform the reduction in section \ref{sec:iib_supergravity_reduced_on_se__5}.
We provide the main result, namely the 5-dimensional action and scalar potential, in section \ref{sec:the_5d_action_and_scalar_potential}.
Then in section \ref{sec:matching_cal_n_4_gauged_supergravity} we match this reduced theory with ${\cal N} = 4$ gauged supergravity, discussing in detail the structure of the bosonic sector. We conclude in section \ref{sec:discussion} with a discussion of our results and on the possible applications in the context of the gauge/gravity correspondence. In two appendices we provide our conventions and give more details on the reduction of the type IIB equations of motion, completing in this way the proof of consistency of our truncation ansatz.

% section introduction (end)

\section{Reducing IIB supergravity on squashed SE$_5$} % (fold)
\label{sec:iib_supergravity_reduced_on_se__5}

Our starting point is type IIB supergravity 
\begin{eqnarray}
\nonumber S_{\rm IIB} \!\!&=&\!\! \frac{1}{2\kappa_{10}^2}\int \left[ R *1 - \frac{1}{2}d\phi\wedge * d \phi - \frac{1}{2}e^{-\phi} H\wedge * H - \frac{1}{2}e^{2\phi} F_1\wedge * F_1 - \frac{1}{2} e^{\phi} F_3\wedge * F_3\right.\\ [2mm]
\label{eq:actionIIB} && \left. \qquad\; - \frac{1}{4} F_5\wedge * F_5 
 -\, \frac{1}{4} (B\wedge dC_2-C_2\wedge dB)\wedge (dC_4+F_5^{\rm flux})\; \right],
\end{eqnarray}
(we will discuss later how to take into account the self-duality of the 5-form) and its 1/4 supersymmetric AdS$_5 \times {\rm SE}_5$ solutions
\begin{equation}
	\begin{array}{l}
		ds^2 \,=\, ds^2({\rm AdS}_5) + ds^2({\rm SE}_5), \\[4mm]
		F_5^{\rm flux}  = (1+*) 2 k\, {\rm vol}({\rm SE}_5),
	\end{array}
\end{equation}
where $k$ specifies the units of flux of the 5-form.
However, we do not want to restrict ourselves to the small fluctuations around these vacua, but perform an off-shell reduction to 5 dimensions that also includes all possible deformations preserving the SE structure of the internal space.
To this end, we will not fix the metric of the residual 5-dimensional space-time and simply propose a reduction ansatz for the 10-dimensional fields in terms of the available structure forms on the compact SE manifolds.
These follow from the very definition of SE spaces and their SU(2) structure group. 

A regular (respectively, quasi-regular) SE manifold $Y$ can be seen as a U(1) fibration over a K\"ahler--Einstein base manifold (respectively, orbifold) $B_{\rm KE}\,$:
\begin{equation}
ds^2(Y) \,=\, ds^2(B_{\rm KE}) + \eta\otimes\eta\,,
\end{equation}
where $\eta$ denotes the globally defined real 1-form dual to the U(1) Reeb Killing vector, which is related to the $R$-symmetry of the associated dual field theories.
All SE manifolds are characterized by 3 globally defined real 2-forms $J^{i}$, which, together with $\eta$, satisfy the algebraic constraints 
\begin{equation}\label{eq:AlgConstr}
J^{i}\wedge J^{j} \,=\,2\,\delta^{ij}\,{\rm vol}(B_{\rm KE})\;,\qquad \qquad \eta\,\lrcorner\, J^{i} = 0\,
\end{equation}
(${\rm vol}(B_{\rm KE})$ denotes the volume form on $B_{\rm KE}$), as well as the differential conditions
\begin{equation}
	\label{SEstructure}
d\eta \,=\, 2J\;,\quad\qquad 	d \Omega = 3 i\, \eta \wedge \Omega\,,
\end{equation}
where, for later convenience, we defined $J \equiv J^{1}$ and $\Omega \equiv J^{2} + i\, J^{3}$. We also have the Hodge duality relations
\begin{equation}\label{eq:*SEforms}
* \eta = {\rm vol}(B_{\rm KE})\;,\quad\qquad *J^{i} \,=\, J^{i}\wedge \eta\,.
\end{equation}
Our ansatz for the dimensional reduction is then constructed by expressing the metric and the various tensor fields of type IIB supergravity in terms of these globally defined forms.

\subsection{The reduction procedure} % (fold)
\label{sub:the_reduction_procedure}

For the reduction of the 10-dimensional metric in the Einstein frame, we follow \cite{MaldMartTach}:
\begin{equation}\label{eq:10dmetric}
ds^2 = e^{-\frac{2}{3}(4U+V)}ds^2(M)\,+\,e^{2U}ds^2(B_{\rm KE})\,+\,e^{2V}(\eta+ A)\otimes (\eta+ A)\,,
\end{equation}
where $U(x)$ and $V(x)$ are scalars and $A(x)$ is a 1--form on $M$, the external 5-dimensional spacetime with Lorentzian signature $(-++++)$. 
Furthermore, we call $x^\mu$ the coordinates on $M$, and $y^m$ the coordinates on $Y$. 
More details about our notations and conventions are reported in Appendix \ref{sec:conventions}. Together the scalars $U$ and $V$ parameterize the ``breathing mode'' and the ``squashing mode'' of the compact manifold: the former is given by $4U+V$ and controls the overall volume, while the latter is $U-V$ and modifies the relative size of the U(1) fibre with respect to the size of the K\"ahler--Einstein base.

While for the dilaton $\phi$ and the Ramond--Ramond axion $C_0$ we assume trivial dependence on the internal coordinates, for the reduction of the other tensor fields of type IIB supergravity we will perform an expansion in the structure forms $\eta$, $J$ and $\Omega$.
Since we would like the reduction ansatz to be gauge covariant and 
to highlight the symmetries of the reduced theory, non-trivial transformation properties have to be assigned to the fields arising from the reduction.
Infinitesimal gauge transformations of the 5-dimensional action fall into two categories: Kalb--Ramond gauge transformations, which follow from the reduction of the 10-dimensional tensor field gauge transformations, like $B \to B + d \Lambda$, and KK gauge transformations.
The latter are the residual gauge transformations induced by reparameterization of the SE fibre coordinate entering into the definition of $\eta$.
The result of this reparameterization on the various 10-dimensional fields can be computed by evaluating the Lie derivative along the isometry vector, which we can identify by its parameter $\omega(x)$: ${\cal L}_\omega = \imath_\omega d + d \imath_\omega$.
When applied to the vielbein associated to the U(1) fibre of the SE manifold, namely $E^9(x,y) = e^{V(x)}(\eta + A)$, we obtain that
\begin{equation}
	\delta_\omega E^9 = e^{V(x)} d \omega(x),
\end{equation}
which has to be interpreted as an action on the 5-dimensional fields $V(x)$ and $A(x)$, so as to have a gauge covariant reduction.
We therefore deduce that $A(x)$ is a gauge field for the KK transformations
\begin{equation}
	\delta_\omega A = d \omega, \qquad \delta_\omega V = 0,
\end{equation}
and for this reason 10-dimensional forms have to be expanded in terms of $\eta + A$ rather than just $\eta\,$.

It is important to point out that while $J$ is invariant under these transformations, $\Omega$ is not:
\begin{equation}
	{\cal L}_\omega \Omega = \imath_\omega d \Omega + d \imath_\omega \Omega = \imath_\omega\left(3i\, \eta \wedge \Omega\right) = 3 i \,\omega\, \Omega.
\end{equation}
Hence the fields associated to $\Omega$ in the expansion will also inherit  non-trivial transformation properties.

As an example we provide the covariant expansion of the 2-form $B$:
\begin{equation}
	\label{eq:Bcov}
	B \,=\, b_2 + b_1 \wedge(\eta + A) + b^{J} J+ {\rm Re}(b^\Omega\,\Omega)\,,
\end{equation}
where $b_p \equiv b_p(x)$ are $p$--forms on $M$ (throughout the paper we omit the 0 subscript for the scalar fields).
The gauge transformations of the 5-dimensional fields are
\begin{equation}
\begin{array}{rclcrcl}
\delta b_2 &=& d\lambda_1 + \lambda_0\, dA,&\phantom{PIPPO}&\delta b^\Omega &=& 3i\omega b^\Omega\,,\\ [3mm]
\delta b_1 &=& d\lambda_0, &&\delta b^{J} &=& 2\lambda_0,
\end{array}
\end{equation}
where the $\lambda_p \equiv \lambda_p(x)$ parameters come from the expansion of the 10-dimensional Kalb--Ramond gauge 1-form
\begin{equation}
\Lambda\,=\, \lambda_1 + \lambda_0 (\eta + A)\,.
\end{equation}

We can also unveil more interesting features of the effective theory by analyzing the expansion of the curvature $H=dB$
\begin{equation}\label{eq:ExpH}
H\,=\, h_3 + h_2\wedge(\eta + A) + h_1^{J}\wedge J + {\rm Re}\big[ h_1^\Omega\wedge\Omega+ h_0^{\Omega}\,\, \Omega\wedge(\eta + A)\big] \,,
\end{equation}
which, recalling (\ref{SEstructure}), leads to the identifications
\begin{equation}
\begin{array}{rclcrcl}
h_3 &=& db_2 - b_1\wedge dA\,,\;\;&&\;\;h_1^\Omega &=& db^\Omega-3iA\,b^\Omega \,\equiv\, Db^\Omega, \\[3mm]
h_2 &=& db_1\,, \;\;&&\;\; h_0^\Omega &=& 3ib^\Omega, \\[3mm]
h_1^J &=& db^J -2b_1 \,\equiv\, Db^J.
\end{array}
\end{equation}
While the 5-dimensional curvatures $h_3$, $h_2$ and $h_1^J$ are gauge invariant, $ h_0^\Omega$ and $h_1^\Omega$ transform as a charged scalar and its covariant derivative, respectively. 
The non-trivial differential relations (\ref{SEstructure}) among the SE forms further give that $h_1^J$ describes a St\"uckelberg coupling between the axion $b^J$ and the one-form $b_1$, the former being a pure gauge under $\lambda_0$.
This structure is common to flux compactifications, where the gauge symmetries induce on the curvatures of the effective theory the structure of a Free Differential Algebra \cite{DallAgata:2005ff,DallAgata:2005mj}.
In the case at hand the fluxes are the flux of the type IIB Ramond--Ramond 5-form, proportional to $k$, and the non-trivial curvature of the internal manifold, captured by (\ref{SEstructure}).

The expansion of the RR 3-form $F_3  = dC_2 - C_0\, dB$ is easy to derive along the lines of the presentation above.
Using (\ref{eq:ExpH}) as a reference and naming $c_p$ and $g_p$ the coefficients in the expansion of $C_2$ and $F_3$ respectively, we get that
\begin{equation}
\begin{array}{rclcrcl}
g_3 &=& dc_2-c_1\wedge dA - C_0(db_2 - b_1\wedge dA),\\ [4mm]
g_2 &=& dc_1-C_0 db_1,&&g_1^\Omega &=& Dc^\Omega -C_0Db^\Omega,\\[4mm]
g_1^J &=& D c^J - C_0 D b^J,&&g_0^\Omega &=& 3i(c^\Omega - C_0 b^\Omega) \,,
\end{array}
\end{equation}
where $D c^J$ and $Dc^\Omega$ read respectively as $D b^J$ and $Db^\Omega$, with the replacement $b\to c$.

% subsection the_reduction_procedure (end)

\subsection{The self-dual 5-form} % (fold)
\label{sub:the_self_dual_5_form}

The expansion of the 5-form 
\begin{equation}
F_5 \;=\;  F_5^{\rm flux}+dC_4 +\frac12\left(B\wedge dC_2 - C_2 \wedge dB\right)
\end{equation} 
follows the same logic adopted for the other forms
\begin{eqnarray}
F_5&=&f_5+f_4\wedge\left(\eta+A\right)+f_3^J\wedge J+ f_2^J\wedge J\wedge\left(\eta+A\right) + {\rm Re}\left[f_3^\Omega\wedge\Omega + f_2^\Omega\wedge\Omega\wedge\left(\eta+A\right)\right] \nonumber\\ [3mm]
&&\!\!+\; f_1\wedge J\wedge J+f_0\,J\wedge J\wedge\left(\eta+A\right),\label{eq:ExpF5}
\end{eqnarray}
and the identification of the various 5-dimensional field-strengths $f_p(x)$ is
\begin{eqnarray}
\nonumber &&f_0=3\,{\rm Im}\big(b^\Omega\,\overline{c^\Omega}\big) \,+\, k\,,\\[4mm]
\nonumber &&f_1=Da +  \frac{1}{2}\big[\, b^JD c^J + {\rm Re}\big(b^\Omega\overline{Dc^\Omega}\,\big)- \,\,b\leftrightarrow c\,\big],\\[4mm]
\nonumber &&f_2^J=da_1^J  +\frac{1}{2}\left[\,b^Jdc_1-b_1\wedge D c^J - \,\,b\leftrightarrow c\,\right], \\[4mm]
% \end{eqnarray} 
% \begin{eqnarray}
\nonumber && f_2^\Omega=Da_1^\Omega+3ia_2^\Omega+\frac{1}{2}\left[b^\Omega dc_1-b_1\wedge Dc^\Omega+3ic^\Omega b_2- \,\,b\leftrightarrow c\right],	\\[4mm]
\label{Expf_p}&& f_3^J = da_2 - 2 a_3 - a_1^J \wedge dA + \frac12\left[ b_2 \wedge D c^J + b^J (dc_2-c_1\wedge  dA) - \,\,b\leftrightarrow c\right],\quad\\[4mm]
\nonumber && f_3^\Omega = D a_2^\Omega - a_1^\Omega\wedge dA + \frac12\left[b_2 \wedge Dc^\Omega + b^\Omega (dc_2-c_1\wedge dA) -\,\,b\leftrightarrow c\right], \\[4mm]
\nonumber && f_4 = d a_3 +\frac12 \left[b_2 \wedge dc_1 -b_1 \wedge (dc_2-c_1\wedge dA) -\,\,b\leftrightarrow c\right],\\[4mm] 
\nonumber && f_5 = f_5^{\rm flux} + d a_4 - a_3 \wedge dA + \frac12 \left[b_2 \wedge (dc_2-c_1\wedge  dA) -\,\,b\leftrightarrow c\right],
\end{eqnarray}
where the terms containing the 5-dimensional $p$-forms $a_p(x)$ come from the expansion of $C_4$. 
We also introduced 0-form and 5-form fluxes parameterized by $k$ and $f_5^{\rm flux}$, and we defined\begin{equation}
Da \equiv da-2a_1^J-kA\;\qquad \hbox{and} \qquad D a_1^\Omega \equiv da_1^\Omega - 3 i A \wedge a_1^\Omega\,.
\end{equation}
The notation $b\leftrightarrow c$ means repetition of the preceding terms within square brackets with $b$ and $c$ exchanged.
However, the above expansion is obviously redundant because the 5-form also has to satisfy the first order, self-duality relation \begin{equation}
* F_5=F_5\,,
\end{equation}
imposing constraints on the expansion of $C_4$, which have not been taken into account yet.
By reducing this equation (the reduction of the Hodge duality operation is reported in the Appendix, cf.~eq.~(\ref{eq:HodgeStars})), we see that the self-duality constraint amounts to the following relations between the forms on $M$ defined in (\ref{Expf_p}):
\begin{equation}\label{eq:Dualityf_p}
\begin{array}{rclcrcl}
f_5 &=& -2\,e^{-\frac{32}{3}U-\frac{8}{3}V}*f_0, &\phantom{PIPPO}&
f_4 &=& 2\,e^{-8U}*f_1, \\ [4mm]
f_3^J &=& -e^{-\frac{4}{3}U - \frac{4}{3}V}*f_2^J,&&
f_3^\Omega &=& -e^{-\frac{4}{3}U - \frac{4}{3}V} *f_2^\Omega\,.
\end{array}
\end{equation}
As a result, some of the 5-dimensional $a_p$ fields introduced in the expansion above should be integrated out and replaced by the dual expressions following from (\ref{eq:Dualityf_p}).
We therefore need to implement this set of constraints while reducing the terms of the action (\ref{eq:actionIIB}) involving the 5-form.
Being $F_5$ self-dual, its kinetic term vanishes on shell\footnote{The problem of obtaining consistent equations of motion from a Lorentz invariant type IIB action has been solved in \cite{IIBcov} by means of a single scalar auxiliary field. 
However, for the purpose of completing the task of reducing type IIB on a SE manifold, the action in \cite{IIBcov} gives no advantage over (\ref{eq:actionIIB}).}, and for this reason we cannot impose the self-duality constraint in the IIB action as it is.
We can, however, proceed along the following lines.

Consider the simplified case of $F_5 = dC_4$, and focus on the duality between the 1- and 4-form field strengths in 5 dimensions, $f_1$ and $f_4$ respectively, also setting $U=V=0$ and $d\eta=0$. 
Using our expansion ansatz, we can integrate over the internal manifold. 
Naming $V_Y = \frac{1}{2}\int_Y J \wedge J \wedge \eta$ the volume of $Y$, the 5-dimensional action reads
\begin{equation}
	S =  -\frac{V_Y}{8\kappa_{10}^2}\int_M \left[\,f_4\wedge *f_4 + 4 f_1\wedge *f_1\, \right]
	\label{Sf4f1}
\end{equation}
and the duality relation between the two is $f_4 = 2 * f_1$. 
We obtain the correct action for the propagating degrees of freedom if we choose to solve the Bianchi identity for $f_1$ in terms of a 0-form potential $a$ (in this simplified case we have $df_1=0\Rightarrow f_1=da$), while we treat $f_4$ as an auxiliary field to be integrated out. 
The self-duality constraint can then be imposed by adding a Lagrange multiplier of the form
\begin{equation}
	S^\prime = -\frac{V_Y}{8\kappa_{10}^2}\int_M 4 \,f_4 \wedge da
	\label{f4da}
\end{equation}
and varying $S + S^\prime$ with respect to $f_4$.
The equations of motion following from $S + S^\prime$ by varying it with respect to $f_4$ and $a$ are 
\begin{equation}
	* f_4 + 2 \,da = 0\quad \hbox{and} \quad d\left[ 2* da + f_4\right] = 0,
\end{equation}
so that together they impose the duality constraint and reproduce the $f_4$ Bianchi identity $df_4 = 0$ as well as the $a$ equation of motion $d*da =0$.
Although substitution of the $f_4$ equation of motion into the action (\ref{Sf4f1}) makes the latter vanish (as it should), the Lagrange multiplier (\ref{f4da}) now becomes the action for the propagating scalar $a$ with a weight which is twice the original one:
\begin{equation}
	S + S^\prime =  -\frac{V_Y}{8\kappa_{10}^2}\int_M 8 \,da \wedge * da\,.
\end{equation}

In order to apply this procedure to the full IIB action we need to take into account also the Chern--Simons couplings in the definition of the $F_5$ curvature.
We then need to reduce 
\begin{eqnarray}
\nonumber S_{ F_5} &=& \frac{1}{2\kappa_{10}^2}\int \left[ - \frac{1}{4} F_5\wedge * F_5   \,-\, \frac{1}{4} (B\wedge dC_2-C_2\wedge dB)\wedge (dC_4+F_5^{\rm flux})\; \right]\\ [2mm]
&=& \frac{1}{2\kappa_{10}^2}\int \left[ - \frac{1}{4} F_5\wedge * F_5   \,-\, \frac{1}{4} L_5\wedge F_5\; \right],
\end{eqnarray}
where we introduced $L_5\equiv B\wedge dC_2-C_2\wedge dB$, which we expand as $F_5$ in (\ref{eq:ExpF5}), with $f_p\to l_p$. Plugging the reduction ansatz in $S_{F_5}$ and integrating over the compact manifold we find
\begin{equation}
\begin{array}{rcl}
S_{ F_5} \!\!\!&=&\!\!\! \displaystyle-\frac{V_Y}{8\kappa_{10}^2}\int_M \left[ e^{\frac{32}{3}U+\frac{8}{3}V} f_5\wedge *f_5 + e^{8U}f_4\wedge *f_4 + 2e^{\frac{4}{3}U+ \frac{4}{3}V}f_3^J\wedge*f_3^J \right.\\ [6mm]
 && + 2e^{\frac{4}{3}U+\frac{4}{3}V}f_3^\Omega\wedge*\overline{f_3^\Omega}+ \;2e^{-\frac{4}{3}U-\frac{4}{3}V} f_2^J\wedge*f_2^J + 2 e^{-\frac{4}{3}U- \frac{4}{3}V}f_2^\Omega\wedge*\overline{f_2^\Omega} \\ [5mm]
 && + 4 e^{-8U}f_1\wedge *f_1 + 4e^{-\frac{32}{3}U-\frac{8}{3}V} f_0^2 *1 + \; 2 l_5 f_0 -2 l_4 \wedge f_1 +2 l_3^J \wedge f_2^J  \\[5mm]
 && \left.+ 2 {\rm Re}(l_3^\Omega\wedge  \overline{f_2^\Omega})- 2l_2^J\wedge  f_3^J -2{\rm Re}(l_2^\Omega\wedge \overline{f_3^\Omega})+ 2l_1\wedge f_4 -2l_0f_5\right].
\end{array}\label{F5action}
\end{equation}
Also in this case we would like the self-duality constraint to be imposed so that the dual degrees of freedom $a_4$, $a_3$ and $a_2$ can be integrated out.
This can be achieved by the procedure outlined above, taking into account that there are Chern--Simons terms to be removed from the curvature definitions.
A further special treatment has to be reserved to the duality relation between $a_2^\Omega$ and $a_1^\Omega$, given by the last equation in (\ref{eq:Dualityf_p}).
Recalling (\ref{Expf_p}), already at first glance it is obvious that this equation cannot be really used to integrate out $a_2^\Omega$, because the latter appears both as a naked potential in $f_2^\Omega$ as well as a curvature in $f_3^\Omega$.
On the other hand, we recall that in 5 dimensions 2-forms can satisfy a ``self-duality'' condition corresponding to a first-order equation of motion (see for instance \cite{PilchTownsendVanN,GZCD}).
A complex self-dual 2-form has the same degrees of freedom as a real 2-form satisfying a second order equation of motion with a mass term.
We should therefore interpret the last of (\ref{eq:Dualityf_p}), rewritten as
\begin{equation}
D a_2^\Omega - a_1^\Omega\wedge dA \,+\,e^{-\frac{4}{3}U - \frac{4}{3}V}* (Da_1^\Omega+3ia_2^\Omega) \,+\; b\;{\rm and }\; c\;{\rm terms}\;=\; 0\,,
\end{equation}
as the first order equation of motion for $a_2^\Omega$, and take this into account when reconstructing the action.
Also, $a_1^\Omega$ are now pure gauge degrees of freedom for $a_2^\Omega$ and should be interpreted as the vector fields eaten up by the tensors in the dualization process \cite{GZCD}.

Motivated by the above arguments, we add the following terms to the original action (\ref{F5action}):
\begin{eqnarray}
\nonumber S' \!\!&=&\! \frac{V_Y}{2\kappa_{10}^2}\int_M \left\{\big(f_5 - \frac{1}{2}l_5\big )k - (f_4 - \frac{1}{2}l_4\big )\wedge Da + (f_3^J + a_1^J \wedge dA - \frac{1}{2}l_3^J\big )\wedge da_1^J \right.\\ [2mm]
&& +\;\left.  {\rm Re}\Big[(f_3^\Omega - Da_2^\Omega +  dA\wedge a_1^\Omega - \frac{1}{2}l_3^\Omega\big )\wedge\big(\overline{Da_1^\Omega+3ia_2^\Omega}\big)\Big]\right\}.
\end{eqnarray}
This set of Lagrange multipliers implements the duality constraints (\ref{eq:Dualityf_p}) upon variation of $S_{F_5}+ S'$ with respect to $f_5,f_4,f_3^J,f_3^\Omega$, the latter being regarded as auxiliary fields. 
By integrating out the auxiliary fields, we eventually obtain the action given in the next section. 

% subsection the_self_dual_5_form (end)

\section{The 5-dimensional action} % (fold)
\label{sec:the_5d_action_and_scalar_potential}

We are now in the position to write down the action of the effective 5-dimensional model.
The reduction of the 10-dimensional Einstein--Hilbert term, as well as the reduction of the 10-dimensional Chern--Simons couplings and of all the kinetic terms, but for $F_5$, are straightforward once one uses the expansions of the various forms given in the previous sections, the reduction of the Ricci scalar following from (\ref{eq:Ricci_ab})--(\ref{eq:Ricci_a9}), and the Hodge duality relations (\ref{eq:HodgeStars}). 
On the other hand, due to the self-duality constraint, for the $F_5$ kinetic term we need to proceed as outlined above.

After completion of the 5-dimensional action, we have verified that it correctly reproduces the 5-dimensional equations of motion following from a direct reduction of the 10-dimensional equations of motion. This proves that our truncation is consistent. 
We discuss the reduction of the 10-dimensional equations and provide the equations of motion for all the 5-dimensional fields in Appendix \ref{sec:5d_equations_of_motion}.

In the following we display the action, organized in three pieces, collecting together the kinetic terms, the 5-dimensional topological couplings and the scalar potential:
\begin{equation}
	S \,=\, S_{\rm kin} + S_{\rm top} +  S_{\rm pot}\,.
	\label{5daction}
\end{equation}
All the definitions of the various curvatures present in this action can be read from the discussion of the reduction ansatz presented in the previous section.

The kinetic terms in 5 dimensions are
\begin{eqnarray}
\nonumber
% \begin{array}{rcl}
S_{\rm kin} \!\!\!\!&=&\! \displaystyle\frac{1}{2\kappa_{5}^2}\int_M \!\Big[ R  *\!1 -\frac{28}{3}dU\wedge *dU -\frac{8}{3}dU\wedge *dV -\frac{4}{3}dV\wedge *dV - \frac{1}{2}e^{\frac{8}{3}U+\frac{8}{3}V}dA\wedge *dA\\ [4mm] 
\nonumber -\!\!\!&\displaystyle\frac{1}{2}&\!\!\!\!\!  d\phi\wedge * d\phi -\frac{1}{2} e^{2\phi}dC_0 \wedge * d C_0- e^{-\frac{4}{3}U-\frac{4}{3}V} f_2^{\Omega} \wedge*\overline{f_2^\Omega} -  e^{-\frac{4}{3}U-\frac{4}{3}V}  f_2^J\wedge * f_2^J - 2e^{-8U} f_1\wedge * f_1\\ [4mm]
\nonumber -\!\!\!&\displaystyle\frac{1}{2}&\!\!\!\! e^{-\phi}   \left(e^{\frac{16}{3}U+\frac{4}{3}V}   h_3\wedge *h_3 + e^{\frac{8}{3}U-\frac{4}{3}V} h_2\wedge *h_2+2e^{-4U}h_1^J\wedge *h_1^J + 2e^{-4U}h_1^\Omega\wedge * \overline{h_1^\Omega}\right) \\[4mm]
\label{Skin}
-\!\!\!&\displaystyle\frac{1}{2}&\!\!\!\! \left.e^{\phi}   \left(e^{\frac{16}{3}U+\frac{4}{3}V}\,g_3\wedge *g_3+e^{\frac{8}{3}U-\frac{4}{3}V}g_2\wedge *g_2+2e^{-4U}g_1^J\wedge *g_1^J + 2e^{-4U}g_1^\Omega\wedge * \overline{g_1^\Omega}\right)\right],
% \end{array}
\end{eqnarray}
where we introduced the 5-dimensional gravitational coupling constant $\kappa_5^2 \equiv \kappa_{10}^2/V_Y$.
The first line comes from the reduction of the 10-dimensional Einstein--Hilbert term, and coincides with the formulae given in \cite{MaldMartTach}.
The first two terms in the second line are the obvious reduction of the 10-dimensional dilaton and axion terms, while the other terms in the same line come from the reduction of the 5-form $F_5$.
The third line arises from the reduction of the $H$ kinetic term and the last line comes from the reduction of the $F_3$ kinetic term.

The 5-dimensional topological couplings read
\begin{eqnarray}
% \begin{array}{rcl}
\nonumber S_{\rm top} \!\!&=& \displaystyle\frac{1}{2\kappa_{5}^2}\int_M \left\{ \frac{i}{3}(\,\overline{Da_1^\Omega + 3ia_2^\Omega}\,) \wedge D(\,Da_1^\Omega +3i a_2^\Omega\,)  + A\wedge da_1^J \wedge da_1^J  \right. \\ [4mm]
\nonumber \!\!\!\!\!\!\!\!&& \displaystyle -\, \frac{1}{2}\,{\rm Re}\left[\big(\,Da_1^\Omega + 3ia_2^\Omega + f_2^\Omega\,\big)\wedge \left( b_2\wedge \overline{Dc^\Omega} + \overline{b^\Omega} (dc_2 -c_1dA) - b \leftrightarrow c \right)\right]\quad \\ [4mm]
\label{SCS}
	 \!\!\!\!\!\!\!\!&& \displaystyle-\,\frac{1}{2}\big(\,da_1^J + f_2^J\,\big)\wedge\big[b_2\wedge Dc^J+b^J\left(dc_2-c_1\wedge dA\right)- \,\,b\leftrightarrow c\,\,\big]\\[4mm]
\nonumber \!\!\!\!\!\!\!\!&&  \displaystyle+\,\frac{1}{2}\left(Da+f_1\right)\wedge\left[b_2\wedge dc_1-b_1\wedge\left(dc_2-c_1\wedge dA\right)-\,\,b\leftrightarrow c\,\,\right]\\ [4mm]
\nonumber	 \!\!\!\!\!\!\!\!&& -\,\frac{1}{2}\,(k+f_0)\left[b_2\wedge \left(dc_2-c_1\wedge dA\right)-\,\,b\leftrightarrow c\,\,\right]\bigg\}
% \end{array}
\end{eqnarray}
and follow from the reduction of the 10-dimensional topological terms (and the Lagrange multipliers necessary to impose the self-duality constraint of the 5-form $F_5$).
Although we put it among the topological terms, we stress once more that $\overline{a_2^\Omega} D a_2^\Omega$ is the kinetic term for the complex 2-form $a_2^\Omega$, which satisfies first order equations.

Finally, the scalar potential terms can be collected in
\begin{equation}
\begin{array}{rcl}
S_{\rm pot} \:=\; \displaystyle\frac{1}{2\kappa_{5}^2}\int_M \big(-2\mathcal V\big) *1 \!\!\!&=&\!\!\! \displaystyle\frac{1}{2\kappa_{5}^2}\int_M \left[ 24\, e^{-\frac{14}{3}U-\frac{2}{3}V} - 4\, e^{-\frac{20}{3}U + \frac{4}{3}V} -2\, e^{-\frac{32}{3}U -\frac{8}{3}V} f_0^2
 \right. \\[6mm]
&&\qquad\qquad\displaystyle\left. - \;e^{-\frac{20}{3}U-\frac{8}{3}V}\left(e^{-\phi}|h_0^\Omega|^2+e^{\phi}|g_0^\Omega|^2\right)  \right]*1\,,
\end{array}
\label{Spot}
\end{equation}
where the first two terms come from the reduction of the 10-dimensional Einstein--Hilbert action, the $f_0$ term comes from the reduction of the 5-form terms, the $h_0^\Omega$ term arises from the reduction of the $H$ kinetic term and the one containing $g_0^\Omega$ comes from the reduction of the $F_3$ kinetic term.

% section the_5d_action_and_scalar_potential (end)
% section iib_supergravity_reduced_on_se__5_ (end)

\section{Matching ${\cal N} = 4$ gauged supergravity} % (fold)
\label{sec:matching_cal_n_4_gauged_supergravity}

In this section we are going to compare the action (\ref{5daction})--(\ref{Spot}) with the one expected for an \hbox{${\cal N} = 4$} gauged supergravity in 5 dimensions, as presented in \cite{SchonWeidner}.
Before starting the match, we are going to justify the claim that the effective action preserves half of the allowed supersymmetries at the lagrangian level (the vacua will further break them).

While so far we focussed only on the bosonic sector, for this discussion we analyze the reduction ansatz for the 10-dimensional gravitino fields.
Type IIB supergravity contains 2 Majorana--Weyl gravitinos of the same chirality $\Psi_M^{\alpha}$, where $\alpha=1,2$, while here $M$ is a 10-dimensional spacetime index.
In order to consistently reduce these fields to 5 dimensions, we employ once again the structure group of the internal manifold, and the fact that it can be associated with the existence of 2 globally defined spinors $\zeta^{1,2}(y)$, being one the charge conjugate of the other \cite{Friedrich}.
We use these internal spinors to expand the 5-dimensional spacetime components of each of the 10-dimensional gravitinos as
\begin{equation}
	\Psi^\alpha_\mu(x,y) = \psi_\mu^{\alpha\,1}(x) \otimes \zeta^1(y) + \psi_\mu^{\alpha\,2}(x) \otimes \zeta^2(y)\,.
\end{equation}
The resulting 5-dimensional gravitinos can then be combined into 4 symplectic-Majorana fermions $\psi_\mu^i$, satisfying
\begin{equation}
	\overline\psi{}_{\mu\,i} \equiv (\psi^i_{\mu})^{\dagger} \gamma^0= \Omega_{ij}(\psi_{\mu}^j)^TC\, ,
\end{equation}
where $\Omega_{ij}$ is the USp(4) invariant symplectic form and $C$ is the charge conjugation matrix.

The above argument is reinforced by noticing that the 5-dimensional fields obtained in our reduction organize in ${\cal N} = 4$ multiplets.
The bosonic spectrum of the truncated theory consists of the metric $g_{\mu\nu}$, 4 vector fields $(A\,,\,a_1^J\,,\,b_1\,,\,c_1)$, 4 tensors $(b_2\,,\,c_2\,,\,a_2^\Omega)$ (recall that $a_1^\Omega$ does not describe degrees of freedom independent of $a_2^\Omega$) and 11 scalars $(U,\,V,\,C_0,\,\phi ,\, a,\, b^J,\, b^\Omega ,\, c^J,\, c^\Omega)$.
The tensor fields in 5-dimensional gauged supergravity arise from the dualization of vector fields.
Hence we can organize these fields in the gravitational multiplet plus two vector-tensor multiplets of 5-dimensional, ${\cal N}=4$ supergravity:
\begin{equation}\nonumber
\begin{array}{rcl}
	&& \!\!\!\{{\rm graviton},\, 6 \,{\rm vectors}, 1 \;{\rm real\;scalar}  \}\\[3mm]
&\!\!\!\!\!2 \;\times&\!\!\! \{{\rm 1\;vector}\,,\,{\rm 5\;real\;scalars} \}\,.
\end{array}
\end{equation}
As a further check we are going to show that the scalar fields are coordinates on the expected coset manifold
\begin{equation}
	{\cal M}_{\rm scal} = {\rm SO}(1,1) \times \frac{{\rm SO}(5,2)}{{\rm SO}(5)\times {\rm SO}(2)}\,,
	\label{coset}
\end{equation}
with the first factor spanned by the scalar in the gravitational multiplet and the second factor parameterized by the scalars in the vector multiplets. We will also show that the vector fields split into $7+1$, seven transforming in the fundamental representation of SO(5,2) $A^M$, and one being a singlet $A^0$.
As mentioned above, the gauging procedure will require the introduction of tensor fields dual to 4 of the vector fields.

\subsection{The ungauged theory} % (fold)
\label{sub:the_ungauged_theory}

In order to properly recognize the couplings of the effective supergravity theory, and to understand which are the contributions following from the gauging procedure, we need to identify the fields in the dimensional reduction with the $\mathcal N=4$ supergravity fields. For this task, we find it convenient to switch off all the gauge interactions and look at the ungauged theory.
This means that we set $k = 0$, i.e.~switch off the Ramond--Ramond 5-form flux, and we take the internal manifold $Y$ to be $K3\times S^1$, so that $d \eta = dJ = d \Omega = 0$.
Then our consistent truncation, preserving the modes associated with $J,\,\Omega,\,\eta\,$, corresponds to a sub-sector (2 vector multiplets only) of the ${\cal N}=4$ ungauged supergravity describing the massless fluctuations around the $\mathbb R^{1,4}\times K3\times S^1$ vacuum.

Our first step is going to be the identification of the scalar $\sigma$-model with ${{\cal M}_{\rm scal}}$ of equation (\ref{coset}). We start by noticing that, once we forget about all the other interactions, the kinetic terms of the scalar fields in (\ref{Skin}) are polynomial in all the fields, but $U$, $V$ and $\phi$.
This suggests that a good identification of the scalar manifold (\ref{coset}) could be obtained via the so-called solvable parameterization, which involves a rewriting of the coset generators in terms of the commuting Cartans and a set of nilpotent generators. 
We then identify the coordinates on ${\cal M}_{\rm scal}$ with the scalars of the dimensional reduction, and match its $G$-invariant metric with the scalar kinetic matrix.

The generators of the $\mathfrak{so}$(5,2) algebra in the fundamental representation are $(t_{MN})_P{}^Q= \delta^Q_{[M}\,\eta^{\phantom{Q}}_{N]P}$, where $M,N,P,Q = \{1,2,\ldots,7\}$ and $\eta = {\rm diag}\{-----++\}$, with commutation relations
\begin{equation}
[t_{MN}, t_{PQ}] \;=\;  \eta_{P[M} t_{N]Q} +  \eta_{Q[N} t_{M]P}\,.
\end{equation}
The $\mathfrak{so}$(5,2) solvable subalgebra is spanned by the two Cartan generators
\begin{equation}
	C_1 = t_{16}, \qquad \qquad C_2 = t_{27}
\end{equation}
and by the nilpotent positive root generators
\begin{equation}
\begin{array}{rcl}
	G_1 \; =\; \frac12 \left(t_{17}-t_{26}-t_{67}-t_{12}\right),\qquad &&\qquad G_2\; =\; \frac12 \left(t_{17}+t_{26}-t_{67}+t_{12}\right), \\[5mm]
	G_3 \; =\; \frac{1}{\sqrt{2}} \left(t_{36}+t_{13}\right),\qquad&&\qquad
	G_4 \; =\; \frac{1}{\sqrt{2}} \left(t_{37}+t_{23}\right), \\[5mm]
	G_5 \; =\; \frac{1}{\sqrt{2}} \left(t_{46}+t_{14}\right),\qquad&&\qquad
	G_6 \; =\; \frac{1}{\sqrt{2}} \left(t_{47}+t_{24}\right), \\[5mm]
	G_7 \; =\; \frac{1}{\sqrt{2}} \left(t_{56}+t_{15}\right),\qquad&&\qquad
	G_8 \;=\; \frac{1}{\sqrt{2}} \left(t_{57}+t_{25}\right).
\end{array}
\end{equation}
Using these generators, we can pick a coset representative 
\begin{equation}
	L \,=\, \left(\prod_{i=0}^7\,e^{x_{8-i} G_{8-i}}\,\right) e^{\phi_2 C_2}e^{\phi_1 C_1}\,
\end{equation}
and define the symmetric matrix
\begin{equation}
	M_{MN} \,=\, \left(L L^T\right)_{MN},
\end{equation}
with inverse $M^{MN}$.
The metric on the ${\cal N}=4$ scalar manifold (\ref{coset}) follows from \cite{SchonWeidner}
\begin{equation}
-\frac 12 ds^2({\cal M}_{\rm scal}) \;=\; -\frac{3}{2}\Sigma^{-2}d\Sigma \otimes d\Sigma + \frac{1}{16} d M_{MN}\otimes d M^{MN},
\end{equation}
where $\Sigma$ is the scalar parameterizing the SO$(1,1)$ factor.
Explicitly we get
\begin{equation}
	\begin{array}{ll}
	ds^2({\cal M}_{\rm scal}) =\; \displaystyle 3\Sigma^{-2}(d\Sigma)^2  
	&\!\!\!\displaystyle + \;\frac{1}{4}\Big[d \phi_1^2 + d \phi_2^2 + \frac12 e^{-\phi_1}(dx_3 - x_2 dx_4)^2 +\frac12 e^{-\phi_1}(dx_5 - x_2 dx_6)^2\\ [5mm]
&\!\!\! \displaystyle+\; \frac12 e^{-\phi_1}(dx_7 - x_2 dx_8)^2+ \frac12 e^{-\phi_2} (dx_4^2+ dx_6^2 + dx_8^2)+ \frac12 e^{\phi_2-\phi_1}dx_2^2 \\[5mm]
	&\!\!\! \displaystyle+\; \frac12 e^{-\phi_1-\phi_2}(dx_1- \frac12 x_3 dx_4 - \frac12 x_5 dx_6 - \frac12 x_7 dx_8)^2\Big].
	\end{array}
\end{equation}
It is then straightforward to see that this expression matches the scalar kinetic terms of our dimensional reduction given in (\ref{Skin}), provided we identify the Cartan coordinates as
\begin{equation}
	\phi_1 = 4U-\phi\;, \qquad \phi_2 = 4U + \phi
\end{equation}
and the nilpotent ones as
\begin{equation}
\begin{array}{rclcrcl}
x_1 &=& 4a + 2c^J b^J + 2 {\rm Re}(b^\Omega \overline{c^\Omega})\, && x_2 &=& 2 C_0\,,\\ [4mm]
x_{3,5,7} &=& 2\sqrt 2\,\{\,{\rm Re}c^\Omega,\,{\rm Im}c^{\Omega},\,c^J\,\}\,,&& x_{4,6,8} &=& 2\sqrt 2\, \{\,{\rm Re}b^\Omega,\,{\rm Im}b^{\Omega},\,b^J\,\}\,.
\end{array}
\label{eq:IdentifScalars}
\end{equation}
Moreover, this forces the identification of the remaining SO$(1,1)$ factor scalar with
\begin{equation}
\Sigma = e^{-\frac{2}{3}(U+V)}.
\end{equation}

Having discussed the scalar manifold, we can now proceed to identify the vector fields in our dimensional reduction with the vectors $\{A^0,A^M\}$ in the ${\cal N}=4$ ungauged supergravity. The general form of the kinetic terms is
\begin{equation}\label{eq:kinvectorsSW}
S_{\rm kin,vec} \;=\; -\frac{1}{2\kappa_5^2}\int_M  \left[\, \Sigma^{-4} \,dA^0\wedge * dA^0 + \Sigma^2 \,M_{MN} dA^M\wedge * dA^N   \,\right],
\end{equation}
where, being the theory ungauged, all vectors have abelian gauge transformations. A study of the reduction of the gauge symmetry associated to the Ramond-Ramond potential $C_4$ (along the lines of subsection \ref{sub:the_reduction_procedure}), shows that $a_1^J$ and $a_1^\Omega$ are not proper gauge fields, and a field redefinition is needed. We find that the correct abelian gauge vectors are given by
\begin{equation}
\tilde a_1^J \; = \; a_1^J  + \frac{1}{2}\left( c^J b_1 - b^J c_1 \right)\;,\qquad \qquad
\tilde a_1^\Omega \; = \; a_1^\Omega + \frac{1}{2}\left( c^\Omega b_1 - b^\Omega c_1 \right) .
\end{equation}
Furthermore, in this ungauged case there is no obstruction to dualize $b_2$ and $c_2$ to 1-forms, which we call respectively $\widehat b_1$ and $\widehat c_1$. 
Implementing all this, the full set of vector kinetic terms provided by the dimensional reduction reads 
\begin{eqnarray}
\nonumber S_{\rm kin,vec} \!\!&=&\!\!  -\frac{1}{4\kappa_5^2}\int_M e^{\frac{8}{3}U+\frac{8}{3}V}dA\wedge *dA + e^{-\frac{4}{3}U -\frac{4}{3}V} \bigg\{  e^{4U-\phi} (db_1)^2 + e^{4U+\phi} (dc_1-C_0db_1)^2  \\ [2mm]
\nonumber &+&\!\!  (d\tilde a_1^J + b^Jdc_1 - c^J db_1)^2 + |d\tilde a_1^\Omega + b^\Omega dc_1 - c^\Omega db_1|^2\\ [3mm]
\nonumber & +&\!\! e^{-4U-\phi} \Big[  d\widehat c_1 -2b^J d\tilde a_1^J - 2{\rm Re}(\,b^\Omega d\overline{\tilde a_1^\Omega}\,) \\ [1mm]
\label{eq:UngaugedVectKin}&+& \big(2a + b^Jc^J + {\rm Re}(b^\Omega\overline{c^\Omega})\big)db_1- \big((b^J)^2 +|b^\Omega|^2\big) dc_1 \Big]^2\\[2mm]
\nonumber & +&\!\! e^{-4U+\phi} \Big[ d\widehat b_1 + C_0d\widehat c_1 + 2(c^J - C_0b^J)d\tilde a_1^J + 2 {\rm Re}\big((c^\Omega - C_0b^\Omega)d\overline{\tilde a_1^\Omega}\big)  \\ [2mm]
\nonumber &+& \Big( 2C_0 a + C_0b^Jc^J + C_0{\rm Re}(b^\Omega \overline{c^\Omega}) -(c^J)^2 - |c^\Omega|^2 \Big)db_1 \\ [2mm]
&-&  \Big(2a - b^Jc^J - {\rm Re} (b^\Omega \overline{c^\Omega}) + C_0(b^J)^2 + C_0|b^\Omega|^2\Big) dc_1\Big]^2 \;\bigg\}*1,\nonumber
\end{eqnarray}
where the squaring of the 2-forms is taken with the $\frac{1}{2}$ factor, e.g. $(db_1)^2 =\frac{1}{2} (db_1)_{\mu\nu} (db_1)^{\mu\nu}$.
Comparing with (\ref{eq:kinvectorsSW}), we derive the identifications
\begin{equation}
\begin{array}{l}
	A = \;\sqrt2\, A^0\,, \qquad\quad b_1 = -A^1-A^6,\qquad 
 c_1 = A^2+A^7,\qquad 
 \tilde a_1^J = A^5\,, \\[3mm]
 \widehat b_1 = -A^1+A^6,\qquad
 \widehat c_1 = A^2-A^7,\qquad
 {\rm Re} \,\tilde a_1^\Omega = A^3,\qquad  {\rm Im}	\, \tilde a_1^\Omega = A^4\,.
\end{array}
\label{eq:IdentifVectors}
\end{equation}
In addition, after dualizing $b_2$ and $c_2$ the topological term agrees with the one expected from $\mathcal N=4$ supergravity.

% subsection the_ungauged_theory (end)

\subsection{The gauged theory} % (fold)
\label{sub:the_gauged_theory}

The last step in our comparison of (\ref{5daction})--(\ref{Spot}) with the expected ${\cal N}= 4$ gauged supergravity action is the computation of the embedding tensor $\Theta$.
This is the tensor specifying how the gauge group is embedded into the duality group and it is all that one needs in order to completely determine the couplings in the lagrangian once the frame of the ungauged theory has been fixed (for a nice review on this approach and further references we suggest \cite{Samtleben:2008pe}).
In the following we show that via $\Theta$ we can match the kinetic terms of the scalar and vector fields, the couplings between the vectors and tensors, as well as the scalar potential.

The explicit form of the embedding tensor can be deduced from the covariant derivative acting on the scalar fields, by looking at the values of the couplings between the scalar and vector fields identified in the previous subsection.
For the case at hand, the general form of the gauged supergravity scalar kinetic term, in the notations of \cite{SchonWeidner}, reads
\begin{equation}\label{eq:GenFormScalKin}
S_{\rm kin,scal}\;=\; \frac{1}{2\kappa_5^2} \int_M  \Big[-3\,\Sigma^{-2}\,d\Sigma \wedge * d\Sigma + \frac{1}{8}\,D M_{MN}\wedge * D M^{MN}\Big],
\end{equation}
where the covariant derivative $D$ has been defined as
\begin{eqnarray}
\nonumber 
D M_{MN} &=& d M_{MN} - 2 A^{\cal P} \big[ X_{{\cal P}} \big]{}_{(M}{}^S M_{N)S} \\ [2mm]
&=& d M_{MN} - 2\big[ A^Pf_P{}^{QR} t_{QR} + A^0 \xi^{PQ} t_{PQ}\big]{}_{(M}{}^S M_{N)S}\label{eq:GenFormCovDer}\\ [2mm]
&\equiv& d M_{MN} + 2 A^Pf_{P(M}{}^Q M_{N)Q}  + 2A^0 \xi_{(M}{}^{Q} M_{N)Q}\,.\nonumber
\end{eqnarray}
Here $X_{{\cal M}} = \Theta_{{\cal M}}{}^\alpha t_{\alpha}$, with $\mathcal M = \{0,M\}$, are the gauge generators, obtained from the product of the embedding tensor $\Theta$ with the $t_{\alpha}$ generators of the duality group SO(1,1) $\times$ SO(5,2).
The tensors $f_{MNP} = f_{[MNP]}$ and $\xi_{MN}= \xi_{[MN]}$ are then the embedding tensor components\footnote{This is actually a special case of the corresponding formula in \cite{SchonWeidner}, in that we are setting to zero the $\xi^M$ components of the embedding tensor therein. 
As it will be clear in the following, this is sufficient to describe the gauged supergravity arising from the dimensional reduction considered in this paper. 
Furthermore, we are reabsorbing the gauge coupling constant $g$ appearing in \cite{SchonWeidner} in the definition of the embedding tensor.} and the indices are raised and lowered with the metric $\eta_{MN}$.

By comparing (\ref{eq:GenFormScalKin}) with the scalar kinetic terms arising from the dimensional reduction, also recalling the field identifications (\ref{eq:IdentifScalars}) and (\ref{eq:IdentifVectors}), we find that the non-vanishing components of the embedding tensor are 
\begin{equation}
\nonumber f_{125} = f_{256} = f_{567} = - f_{157} = -2,
\end{equation}
\begin{equation}\label{eq:OurEmbTensor}
\xi_{34}= -3\sqrt 2,\qquad\qquad \xi_{12}=\xi_{17}=-\xi_{26}= \xi_{67} = -\sqrt 2\, k\,,
\end{equation}
together with the ones given by cyclic permutations of the indices.
The higher-dimensional origin of $f_{MNP}$ resides in the geometric flux associated with the non-closure of $\eta$, \hbox{$d\eta = 2J$}, while $\xi_{34}$ arises from the geometric flux $d\Omega = 3i\,\eta\wedge \Omega$. The remaining non-zero $\xi_{MN}$ parameters derive from the Ramond-Ramond 5-form flux described by $k$.

Since the embedding tensor specifies how the gauge group $G$ is embedded into the duality group, we can now discuss the interesting features of the gauge group of the theory at hand.
As we have already noticed, after the gauging procedure we are left with four gauge vector fields, sitting in the adjoint of $G$.
We name $(t_\Lambda)_M{}^N$ the corresponding gauge generators, where $\Lambda=0,1,2,3$, and $A^\Lambda= \{A, b_1,c_1,\tilde a_1^J\}$ the vector fields to which they couple, so that the gauge covariant derivative reads $D = d- A^\Lambda t_{\Lambda}$.
By direct comparison with (\ref{eq:GenFormCovDer}) we find that
\begin{equation}\label{eq:GaugeGen}
t_0 = -6\,t_{34}+ 4k\,G_1\,, \qquad t_1= 4\sqrt 2\,G_8\,,\qquad t_2 = 4\sqrt 2\,G_7\,,\qquad t_3= 8\,G_1\,,
\end{equation}
the only non-trivial commutator being $[t_1,t_2]=-2t_3\,$.
The resulting gauge group $G$ is then a product of the 3-dimensional Heisenberg group with a U(1) subgroup of the USp(4) $\simeq$ SO(5) $R$-symmetry\footnote{As a check, we verified that the associated 5-dimensional gauge transformations match the ones derived by dimensionally reducing the gauge symmetry of the 10-dimensional forms and the diffeomorphisms reparameterizing the U(1) fibre of the internal manifold (cf. subsection \ref{sub:the_reduction_procedure}).}
\begin{equation}
	G = {\rm Heis}_3 \times {\rm U(1)}_R\,,
\end{equation}
where the U(1)$_R$ factor is generated by $t_{34}$. From (\ref{eq:GaugeGen}) we see that the vector fields that are not in the adjoint representation of $G$ also transform non-trivially under the action of the gauge generators.
Actually, we can split the indices of the various vector fields according to their transformation properties.
We take the original set $\{A^0, A^M\}$ and split it into the $A^\Lambda$ vectors in the adjoint of $G$, and the $A^I$ vectors, in a non-trivial representation of $G$.
With this choice of basis, we can rewrite the 4 gauge generators as
\begin{equation}
	t_\Lambda = \left(\begin{array}{cc}
	-f_{\Lambda \Sigma}{}^\Gamma & (t_\Lambda)_\Sigma{}^I \\[2mm]
	0 & (t_\Lambda)_J{}^I
	\end{array}\right),
\end{equation}
where $f_{\Lambda \Sigma}{}^\Gamma$ are the structure constants of the gauge group in the adjoint representation, $(t_\Lambda)_J{}^I$ are the generators under which the vector fields are in a symplectic representation and $(t_\Lambda)_\Sigma{}^I$ are the generators of the gauge group that mix the vector fields in the adjoint with the ones that are going to be dualized to tensor fields.
This is the (so far known) most general structure of the couplings between vector fields and tensor fields in 5 dimensions, compatible with supersymmetry \cite{Bergshoeff:2002qk}.
For the case at hand, the only adjoint structure constant is \hbox{$f_{{1}{2}}{}^{3}= - f_{{2}{1}}{}^{3} = -2$}. This implies that $(t_3)_{\Sigma}{}^\Gamma$ vanishes, 
in agreement with the fact that, while the vector fields are in a faithful representation of the gauge group, the adjoint representation gets rid of all the abelian ideals in the non-abelian factors (see \cite{Dall'Agata:2007sr} for a general discussion of this mechanism for flux compactifications).
The only non-trivial $(t_\Lambda)_J{}^I$ is given by $t_0$, the U(1) generator under which $\tilde a_1^\Omega$ and its dual $\tilde a_2^\Omega$ are charged, while the $(t_\Lambda)_\Sigma{}^I$ generators deserve a further discussion, because to our knowledge this is the first realization of such a structure in a stringy reduction without maximal supersymmetry. The introduction of tensor fields is motivated by the fact that whenever there are vector fields transforming under a non-adjoint representation of the gauge group, their field strengths do not generically transform covariantly under gauge transformations, and one needs to employ tensor field transformations to close the Jacobi identities and make the generalized field strengths covariant \cite{DallAgata:2005ff,deWit:2005ub}.
For ${\cal N}=4$, 5-dimensional gauged supergravity, the covariant field strengths of the vector fields include 2-forms $B_{\mathcal N}$ and read \cite{SchonWeidner}
\begin{equation}
\mathcal H^{\mathcal M} = dA^{\mathcal M} + \frac12\,  X_{\mathcal N\mathcal P}{}^{\mathcal M}\, A^{\mathcal N}\wedge A^{\mathcal P} + Z^{\mathcal M \mathcal N} \,B_{\mathcal N},
\end{equation}
where $X_{{\cal M}}$ are related to the embedding tensor as in (\ref{eq:GenFormCovDer}) and $Z^{ M  N}= \frac{1}{2}\xi^{MN}$ collects  the tensor couplings (the $Z^{0M}$ components vanish in our case).
Employing (\ref{eq:OurEmbTensor}), we see that for our reduction the gauge group $G$ is represented on the gauge fields by the following curvatures in the adjoint representation:
\begin{equation}\label{FirstFieldStr}
\begin{array}{rcl}
\mathcal H^0 &=& dA^0,\\ [3mm]
\mathcal H^1+ \mathcal H^6 &=& d(A^1+A^6),\\ [3mm]
\mathcal H^2+ \mathcal H^7 &=& d(A^2+A^7),\\ [3mm]
\mathcal H^5 &=& dA^5 - 2(A^1+A^6)\wedge(A^2+A^7),
\end{array}
\end{equation}
where ${\cal H}^0$ is the field strength of the U(1)$_R$, while the other 3 curvatures are the 3-dimensional realization of the Heisenberg group. The field strengths of the U(1)$_R$-charged vectors, naturally combining in a complex field, follow from the embedding tensor above and read
\begin{equation}
\begin{array}{rcl}
\mathcal H^3 +i \mathcal H^4&=& \displaystyle d(A^3 +i A^4) - 3i\sqrt 2 A^0\wedge (A^3+iA^4) +{{\frac{3i}{\sqrt 2}}}  (B_3+iB_4).
\end{array}
\end{equation}
Recalling (\ref{eq:IdentifVectors}), we identify this expression with the combination $D\tilde a_1^\Omega +3i \tilde a_2^\Omega$ derived from the dimensional reduction, where again a suitable field redefinition $\tilde a_2^\Omega = a_2^\Omega +\frac{1}{2}(c^\Omega b_2 -b^\Omega c_2)$ was required. Finally, the most interesting couplings arise in the curvatures for the vector fields in a non-trivial representation of the Heisenberg group:
\begin{equation}\label{LastFieldStr}
\begin{array}{rcl}
\mathcal H^1-\mathcal H^6 &=& \displaystyle d(A^1-A^6) + (4 A^5+2\sqrt 2 k\, A^0) \wedge (A^2+A^7) - \sqrt 2 k \,(B_2 - B_7),\\ [4mm]
\mathcal H^2-\mathcal H^7 &=& \displaystyle d(A^2-A^7) - (4 A^5+2\sqrt 2 k\, A^0) \wedge (A^1+A^6) + \sqrt 2 k \,(B_1 - B_6).
\end{array}
\end{equation}
The corresponding vector combinations do in fact transform under $t_3$, as well as under the gauge transformations of the tensor fields:
\begin{equation}
\begin{array}{rcl}
\delta (A^1-A^6)&=& d(\Lambda^1-\Lambda^6) - 4 (A^2+A^7)\Lambda^5 + (4 A^5+ 2\sqrt 2k\,A^0)(\Lambda^2+\Lambda^7) + \sqrt 2 k \, (\Xi_2 - \Xi_7), \\ [4mm]
\delta (A^2-A^7)&=& d(\Lambda^2-\Lambda^7) + 4 (A^1+A^6) \Lambda^5 - (4 A^5+ 2\sqrt 2k\,A^0)(\Lambda^1+\Lambda^6) - \sqrt 2 k\, (\Xi_1 - \Xi_6).
\end{array}
\end{equation}
Here, $\Lambda^\mathcal M$ and $\Xi_{\mathcal M}$ denote the gauge tranformations of the vector and tensor fields respectively. In the approach of \cite{SchonWeidner}, the 2-forms $B_1 - B_6$ and $B_2 - B_7$ are dual to the vectors $A_1 - A_6$ and $A_2 - A_7$, and both are kept in the gauged supergravity lagrangian, though some degrees of freedom are not dynamical. 
The duality relation between the respective covariant field strengths arises as the equation of motion for the tensors, as already mentioned in subsection \ref{sub:the_self_dual_5_form} when we discussed the relation between $a_2^\Omega$ and $a_1^\Omega$. 
Now, while we identify $B_1 - B_6$ and $B_2 - B_7$ with the 2-forms $b_2$ and $c_2$, their dual vectors are not directly obtained from the dimensional reduction procedure. 
When discussing the ungauged theory in subsection \ref{sub:the_ungauged_theory}, the latter were introduced by dualizing $b_2$ and $c_2$ to $\widehat b_1$ and $\widehat c_1$, but once we switch on the RR five-form flux $k$ this is not possible any more. 
Indeed, it can be seen that in our 5-dimensional action one can perform suitable partial integrations and cover $b_2$ and $c_2$ with a derivative everywhere, but on the topological term $k (c_2 \wedge db_2 - b_2 \wedge dc_2)$ in (\ref{SCS}), where either $b_2$ or $c_2$ necessarily appear naked. 
This means that the obstruction against dualizing $b_2$ and $c_2$ to vectors is precisely the flux $k$.\footnote{This can also be seen at the 10-dimensional level: without 5-form flux we would be allowed to rewrite the Chern-Simons term in the IIB action (\ref{eq:actionIIB}) as $dB \wedge dC_2 \wedge C_4$, i.e.~neither $B$ nor $C_2$ would ever appear naked into the action.} 
However, we can make contact with the formalism of \cite{SchonWeidner} by noticing that the second order equations of motion for $b_2$ and $c_2$ derived in appendix \ref{sec:5d_equations_of_motion} have the form of a total derivative, and can then be interpreted as first order equations stating the duality relation. 
Indeed, it turns out that the equations for $b_2$ and $c_2$ (see (\ref{eq:b2eq}) and (\ref{eq:Transforms})) can be written as
\begin{equation}\label{eq:SecondOrderEqbc}
d\left[ \Big(m * d-2k\,\omega\Big)\left(\begin{array}{c} b_2 \\ c_2\end{array}\right)  + {\rm other}\;{\rm terms} \right]\;=\;0\,,
\end{equation}
where we introduced the matrices
\begin{equation}
m= e^{\frac{16}{3}U+\frac{4}{3}V+\phi} \left(\begin{array}{cc} C_0^2 + e^{-2\phi} & -C_0 \\ -C_0 & 1 \end{array}\right)\;,   \qquad\qquad \omega = \left(\begin{array}{cc} 0 & 1\\ -1 & 0 \end{array}\right)\,,
\end{equation}
$m$ being ($e^{\frac{16}{3}U+\frac{4}{3}V}$ times) the SL$(2,\mathbb R)$ analog of the matrix $M$ defined in subsection \ref{sub:the_ungauged_theory}. 
Then we deduce the first order equations
\begin{equation}\label{eq:FirstOrderEqbc}
\Big(m * d-2k\,\omega\Big)\left(\begin{array}{c} b_2 \\ c_2\end{array}\right)  + {\rm other}\;{\rm terms} \;=\; d\left(\begin{array}{c} \widehat b_1 \\ \widehat c_1\end{array}\right).
\end{equation}
These can be used to replace $db_2$ and $dc_2$ in the dimensionally reduced action, and obtain an action consistent with the one of \cite{SchonWeidner}. 
In particular, the terms $h_3\wedge* h_3$ and $g_3\wedge* g_3$ in (\ref{Skin}) become kinetic terms for the vectors $\widehat b_1$ and $\widehat c_1$, and we retrieve the full set of vector kinetic terms (\ref{eq:UngaugedVectKin}) previously derived for the ungauged theory, now covariantized by means of (\ref{FirstFieldStr})--(\ref{LastFieldStr}). 
The physical degrees of freedom propagated by the fields $b_2,\,c_2,\,\widehat b_1\,,\widehat c_1$ are best seen by using the second of (\ref{eq:FirstOrderEqbc}) to eliminate $dc_2$ from the first of (\ref{eq:SecondOrderEqbc}): this yields a Proca equation for a 2-form, with a (scalar dependent) mass term proportional to $k^2$.

We conclude this section by showing that the embedding tensor derived above also reproduces the scalar potential obtained in (\ref{Spot}), via  eq.~(3.16) of \cite{SchonWeidner}. 
It is useful to split the scalar potential of \cite{SchonWeidner} in three addends according to the powers of $\Sigma$, so that we can compare the resulting expressions with those in (\ref{Spot}) according to the powers of $U$ and $V$.
An explicit computation shows that the three resulting pieces are
\begin{eqnarray}
\nonumber &&\!\!\!{\textstyle{\frac{1}{4}}}\,f_{MNP} \,f_{QRS}\, \Sigma^{-2}\big({\textstyle{\frac{1}{12}}}M^{MQ}M^{NR}M^{PS}-{\textstyle{\frac{1}{4}}} M^{MQ}\eta^{NR}\eta^{PS} + {\textstyle{\frac{1}{6}}}\eta^{MQ}\eta^{NR}\eta^{PS}\big) \,\;=\;\, 2\, e^{-\frac{20}{3}U + \frac{4}{3}V},\\ [4mm]
\nonumber &&\!\!\!{\textstyle{\frac{1}{16}}} \,\xi_{MN}\,\xi_{PQ}\,\Sigma^4 \big( M^{MP}M^{NQ} -\eta^{MP}\eta^{NQ} \big) \;\,=\\ [2mm]
\nonumber &&\!\!\!\quad =\; {\textstyle{\frac{9}{2}}} \,{e}^{-\frac{20}{3}U-\frac{8}{3}V -\phi}|b^\Omega|^2 + {\textstyle{\frac{9}{2}}} \,{e}^{-\frac{20}{3}U-\frac{8}{3}V+ \phi}|c^\Omega - C_0 b^\Omega|^2 + {e}^{-\frac{32}{3}U-\frac{8}{3}V} \big[3\,{\rm Im}\big(b^\Omega\,\overline{c^\Omega}\big) + k\big]^2,\\ [4mm]
&&\!\!\!{\textstyle{\frac{1}{12}}}\sqrt 2 \,f_{MNP}\, \xi_{QR} \,\Sigma\, M^{MNPQR} \,\;=\;\, -12\, {e}^{-\frac{14}{3}U-\frac{2}{3}V},
\end{eqnarray}
whose sum is precisely the scalar potential $\mathcal V$ given in (\ref{Spot}).

% subsection the_gauged_theory (end)

% section matching_cal_n_4_gauged_supergravity (end)

\section{Discussion} % (fold)
\label{sec:discussion}

The 5-dimensional supergravity model we have detailed in the previous sections describes the physics of type IIB supergravity compactified on a squashed SE manifold.
In particular, the scalar potential ${\cal V}$ governs the deformations of the internal manifold and the vacuum expectation values of the 10-dimensional form fields. 
We rewrite here its expression (\ref{Spot}) in a more explicit fashion:
\begin{equation}\label{eq:ScalPot}
\begin{array}{rcl}
\mathcal V &=&\displaystyle  -\,12 \,{ e}^{-\frac{14}{3}U-\frac{2}{3}V} +2\, { e}^{-\frac{20}{3}U + \frac{4}{3}V} + \frac{9}{2}\, { e}^{-\frac{20}{3}U-\frac{8}{3}V -\phi}|b^\Omega|^2 \\[4mm] 
 &&\displaystyle+\, \frac{9}{2}\, { e}^{-\frac{20}{3}U-\frac{8}{3}V+ \phi}|c^\Omega - C_0 b^\Omega|^2 + \;{ e}^{-\frac{32}{3}U-\frac{8}{3}V} \big[3\,{\rm Im}\big(b^\Omega\,\overline{c^\Omega}\big) + k\big]^2\,.
\end{array}
\end{equation}
The fact that $\mathcal V$ does not depend on the scalars $a, \,b^J$ and $c^J$ was expected, since the latter play the role of St\"uckelberg fields for some of the vectors. 

Notice that our $\mathcal V$ is an extension of the scalar potential presented in \cite{MaldMartTach}, which is recovered by taking $b^\Omega = c^\Omega = 0$, and choosing the flux parameter $k^2=4$.

We find that the potential (\ref{eq:ScalPot}) has two extrema, which, for $k=2$, are located at
\begin{equation}
U \,=\, V \,=\, b^\Omega \,=\, c^\Omega \,=\, 0\,,\quad{\rm with}\;{\rm arbitrary}\; \phi\;{\rm and} \;C_0\,,
\label{susyvacuum}
\end{equation}
and at
\begin{equation}
e^{4U}=e^{-4V}=\frac{2}{3}\,,\qquad  b^\Omega=\frac{e^{i \theta+\phi/2}}{\sqrt{3}}\,\,,\qquad c^\Omega=b^\Omega  \tau\,, \qquad \tau \equiv (C_0  + i\,e^{-\phi})\,,
\label{nonsusyvacuum}
\end{equation}
where we have 3 flat directions, parameterized by $\phi$, $C_0$ and $\theta$.
Both vacua have a negative value of the cosmological constant $\Lambda \equiv \langle \mathcal V \rangle$ and therefore correspond to anti-de Sitter vacua.
The first one has $\Lambda = -6$, while $\Lambda = - \frac{27}{4}\,$ for the second one. 

The first extremum is supersymmetric for any value of $k$ and, having $U = V$, is associated with the round metric (by analogy with the case where the SE manifold is $S^5$). 
The second extremum has $U\neq V$, instead.
Hence the internal metric is squashed and non-Einstein. 
From the higher-dimensional viewpoint, it corresponds to a non-supersymmetric solution of type IIB supergravity found in \cite{RomansIIBsols,GunaydinRomansWarner}, which is the 5-dimensional analog of the Pope--Warner solution in 4 dimensions \cite{Pope:1984bd}. 
From the 5-dimensional point of view, this squashed vacuum is identified with the  SU(3)$\times$U(1) invariant vacuum of the gauged SO(6) maximal supergravity in 5 dimensions, derived in \cite{GunaydinRomansWarner}, as we will justify shortly.

The masses of the scalar fluctuations around the supersymmetric and non-supersymmetric vacua (\ref{susyvacuum}) and (\ref{nonsusyvacuum}) are obtained by computing the eigenvalues of the mass matrix
\begin{equation}
	M^i{}_j = 2\, K^{ik}\, \partial_j \partial_k {\cal V}
\end{equation}
evaluated at the critical points. 
Here, the index $i$ runs over the  scalar fields appearing in $\mathcal V$, collected in the array $\varphi^i \equiv \{U,V,{\rm Re}\, b^\Omega,{\rm Im}\, b^\Omega,{\rm Re}\, c^\Omega,{\rm Im}\, c^\Omega,\phi,C_0\}$, while the matrix $K_{ij}$ is the kinetic matrix of the normalized scalar fields
\begin{equation}
	-\frac12 K_{ij}(\varphi)\, \partial_\mu \varphi^i \partial^\mu \varphi^j.
\end{equation}

\subsection{The supersymmetric vacuum} % (fold)
\label{sub:the_supersymmetric_vacuum}

At the supersymmetric critical point we find that the mass eigenstates of the field fluctuations can be collected in the following table, where angle brackets denote the choice of vacuum expectation values for the dilaton and axion moduli:

\definecolor{gray}{rgb}{.9,.9,.9}

\renewcommand{\arraystretch}{1.5}

\begin{equation}
	\begin{array}{ccc}\hline
\rowcolor{gray}	{\rm Mass \ Eigenstate} &\phantom{sp}& m^2\\\hline
	 4 \delta U +\delta V && 32  \\
\rowcolor{gray}		\delta U-\delta V && 12\\
\;\langle C_0 e^{2 \phi}\rangle {\rm Re}\,\delta c^\Omega - \left(1 + 
\langle C_0^2 e^{2 \phi}\rangle\right) {\rm Re}\,\delta b^\Omega+ \langle e^\phi\rangle {\rm Im}\,\delta c^\Omega && 21\\
\rowcolor{gray}	-\langle C_0 e^{2 \phi}\rangle {\rm Im}\,\delta c^\Omega + \left(1 +\langle C_0^2  e^{2 \phi}\rangle\right) {\rm Im}\,\delta b^\Omega + \langle e^\phi\rangle {\rm Re}\,\delta c^\Omega && 21\\
\;\langle C_0 e^{2 \phi}\rangle {\rm Im}\,\delta c^\Omega - \left(1 + 
\langle C_0^2 e^{2 \phi}\rangle\right) {\rm Im}\,\delta b^\Omega + \langle e^\phi\rangle {\rm Re}\,\delta c^\Omega && -3\\
\rowcolor{gray}	-\langle C_0 e^{2 \phi}\rangle {\rm Re}\,\delta c^\Omega + \left(1 +\langle C_0^2 e^{2 \phi}\rangle\right) {\rm Re}\,\delta b^\Omega + \langle e^\phi\rangle {\rm Im}\,\delta c^\Omega && -3\\
	\delta \phi && 0\\
\rowcolor{gray}		\delta C_0 && 0\\\hline
	\end{array}
\end{equation}

\medskip

\noindent
Although this expansion is general and valid for any internal SE manifold, in the following we specialize our analysis to the case of the 5-sphere, so that we have some direct control on the dual field theory.
However, the gauge/gravity correspondence relations we derive in this way are valid in general for any ${\cal N} = 1$ superconformal field theory in 4 dimensions.

The first step is the identification of the linear combinations in the table above with the appropriate states in the spectrum of KK modes in the expansion around the 5-sphere vacuum \cite{KRV}.
The ansatz we have chosen for the type IIB metric and tensor fields is compatible with a truncation of the $S^5$ spectrum to SU(3) singlets in the decomposition
\begin{equation}
	{\rm SO}(6) \simeq {\rm SU}(4) \to {\rm SU}(3) \times {\rm U}(1).
\end{equation}
This is indeed the type of truncation that follows by requiring that an SU(2) structure group is preserved on $S^5$.
This truncation leaves us with an $\mathcal N=2$ spectrum, which could also be obtained by retaining the states which are left-invariant forms in the reduction on\footnote{The standard parameterization of the 5-sphere $S^5 = $ SO(6)/SO(5) leads to the non-supersymmetric truncation keeping only the breathing mode. In both cases, the consistency of the truncation follows from arguments parallel to the ones applied in \cite{ExploitingN=2}.} 
\begin{equation}
	S^5 = \frac{\rm SU(3)}{\rm SU(2)},
\end{equation} 
where the SU(2) structure group is identified with the denominator of the coset.
By inspection of the SU(4) representations of the spectrum in \cite{KRV}, we see that only states in the singleton, massless graviton and in the first two KK iterations can survive the truncation to singlets of SU(3) $\subset$ SU(4). 
This indeed reorganizes the spectrum of fluctuations around the supersymmetric vacuum in ${\cal N} = 2$ multiplets as follows\footnote{We refer to \cite{T11} for nomenclature and for the structure of $\mathcal N=2$ multiplets.}: the gravity multiplet $(g_{\mu\nu}\,,\,2\psi_\mu \,,\, A_\mu)$ and a hypermultiplet $(\chi, 4 \varphi)$ from the massless ${\cal N} = 8$ graviton; a semi-long massive gravitino multiplet $(2\tilde\psi_\mu\,,2A_\mu, \, 2b_{\mu\nu},4\chi)$ from the first KK iteration; a long vector multiplet $(B_\mu\,,4 \chi\,, 4 \varphi)$ from the second KK iteration.
This multiplet structure arises from the ${\cal N}=4$ massless multiplets of sections \ref{sec:iib_supergravity_reduced_on_se__5} and \ref{sec:matching_cal_n_4_gauged_supergravity} via a spontaneous gauge and partial supersymmetry breaking mechanism at the vacuum.
From the spectrum of the vector fields at this vacuum we can see that out of the 4 original gauge vector bosons only one, the graviphoton associated to the U(1)$_R$ symmetry, is massless, while the other 3 vector fields have a mass, breaking completely the 3-dimensional Heisenberg group:
\begin{equation}
	G = {\rm Heis}_3 \times {\rm U}(1)_R \to {\rm U}(1)_R\,.
\end{equation}
This is easily checked by looking at the quadratic couplings of the vector fields in the kinetic terms of the scalars, after having canonically normalized the vector kinetic terms.
In detail, we find that the linear combination $\tilde a_1^J + A$ has mass $m^2 = 24$, $b_1$ and $c_1$ have $m^2 = 8$, while the combination $2\,\tilde a_1^J-A$ remains massless.
By expanding the covariant derivative $d - A^\Lambda t_\Lambda$ we can also realize that the latter combination is associated to $2\, t_{34}$, which is the surviving U(1)$_R$ gauge symmetry generator.
In addition, partial supersymmetry breaking gives mass to half of the gravitino fields, which end up in the massive gravitino multiplet.
All the scalars having St\"uckelberg couplings to some of the vectors are removed from the analysis of our spectrum, being simply regarded as the longitudinal degrees of freedom of those vectors.

For what concerns the scalar fields appearing in the scalar potential, we can see that the last 4 states in the table above are part of the ${\cal N} = 8$ gravity multiplet and, after our ${\cal N} = 2$ truncation, they fill a hypermultiplet.
The other states are part of massive KK iterations instead.
The $m^2 = 21$ states can be identified with scalars in the first massive KK tower expansion of the SL$(2,\mathbb R)$-covariant 2-form given by the complex combination of $B$ and $C_2$, while the $m^2 = 12$ state is the squashing mode and the $m^2 = 32$ state is the breathing mode of the SE internal manifold, both sitting in the second KK tower.

In the following we discuss the dual operators related to the various multiplets and show the match of the conformal dimensions with the masses of the scalar fields described in the table above.
For notation and more details on the superfield description of such theories we follow \cite{T11} where a complete analysis of the spectrum of the $T^{1,1}$ manifold and of the dual conformal theories has been presented.
Our analysis is obviously a subcase of the one presented there and yet it is, at the same time, more general, because it is valid for any SE manifold and therefore for any ${\cal N} = 1$ conformal theory.

The ${\cal N} = 2$ massless graviton multiplet corresponds to the stress-energy tensor of the dual gauge theory and can be described by an operator whose lowest component is the stress-energy tensor for the gauge fields 
\begin{equation}
	J_{\alpha \dot \alpha}={\rm Tr}\,\left( W_\alpha \overline W_{\!\dot \alpha}\right) + \ldots,
\end{equation}
where $W_\alpha$ describes the gauge multiplet in the superfield language.
Although what replaces the dots, completing the explicit form of the operator, depends on the details of the theory at hand and especially on the structure of the matter multiplets, which in turn depend on the geometry of the internal SE manifold, we see that our reduction captures the universal part of it, in which only the gauge superfield appears.
The massless hypermultiplet corresponds to the chiral operator
\begin{equation}
	\Phi = {\rm Tr}\, \left(W_\alpha W^\alpha\right) + \ldots,
\end{equation}
or, better, to a linear combination of this operator and the superpotential being orthogonal to the derivatives of the Konishi multiplet.
The conformal dimension of $\Phi$ is $\Delta = 3$ and the masses of the associated scalar fields follow from the general gauge/gravity duality relation between the mass of a scalar field $\varphi$ and its dual operator ${\cal O}_\varphi$:
\begin{equation}
	L^2m_\varphi^2 = \Delta_{{\cal O}_\varphi}(\Delta_{{\cal O}_\varphi}-4),
	\label{ggrelation}
\end{equation}
where $L^2 = 6/|\Lambda|$ is the AdS radius. In this case we have two states with $m^2 = -3$ and two massless states coming from the descendants with $\Delta = \Delta_\Phi +1$.
Both the $J_{\alpha \dot \alpha}$ and $\Phi$ multiplets also appear in the truncations of the massless spectrum of ${\cal N} = 8$ supergravity and indeed they are the product of two singleton fields $W_\alpha\,$.
The first KK iteration follows by taking the product of three singleton fields. The only possible operator built in this way corresponds to our massive gravitino multiplet, and reads
\begin{equation}
	L_{\dot \alpha} ={\rm Tr} \,\left(\overline W_{\!\dot \alpha} W_{\beta}W^\beta\right)+ \ldots. 
\end{equation}
This superfield has conformal dimension $\Delta_L = 9/2$ and contains no scalar components.

Finally, at the second iteration level we have
\begin{equation}
	Q = {\rm Tr}\, \left(W^2 \overline W{}^2\right) + \ldots,
\end{equation}
which has $\Delta_Q = 6$ and corresponds to the long vector multiplet.
This operator contains 4 scalar fields associated with the conformal dimensions $\Delta = \Delta_Q$, twice $\Delta = \Delta_Q+1$ and $\Delta = \Delta_Q+2$.
The obvious dual massive states have $m^2 = 12$, twice $m^2 = 21$ and $m^2 = 32$.
While the operator $\Phi$ describes a relevant deformation of the gauge theory, $Q$ and $L$ are irrelevant.
We notice that additional states survive the truncation in the expansion of the singleton multiplet, which in this case is simply $W_\alpha\,$.
These states, however, are pure gauge states from the 5-dimensional point of view, corresponding to the fact that ${\rm Tr}\, W_\alpha = 0\,$, being the $W_\alpha$ superfield in the adjoint representation of SU($N$).

% subsection the_supersymmetric_vacuum (end)

\subsection{The susy breaking vacuum} % (fold)
\label{sub:the_susy_breaking_vacuum}

At the non-supersymmetric critical point we find that the mass eigenstates are complicated combinations depending on the expectation values of the axio-dilaton and the $\theta$ parameter. 
The masses however do not depend on these values and for the flux choice $k=2$ read
\begin{equation}\label{eq:NonSusyScalarMasses}
	m^2 = \{\,36,\,36,\,27,\,27,\,9,\,0,\,0,\,0\,\},
\end{equation}
where one of the $m^2 = 36$ states is given by the combination $\delta U+\delta V$.
For the special point of the parameter space $\langle\theta\rangle = 0$, $\langle C_0\rangle = 0$ and $\langle\phi\rangle = 0$, the eigenstates corresponding to the above eigenvalues can be collected in the following table, where we also specified the conformal dimension of the dual operators according to relation (\ref{ggrelation}):\footnote{{\bf Note added in v2}: In agreement with a remark appeared in \cite{SkenderisTaylorTsimpis}, we added the missing $L^2$ factor in the formula relating $m^2$ and $\Delta$, which yields a modification in our previous values of the conformal dimensions. We also fixed a typo in the eigenstate associated with $m^2=9$.}
\begin{equation}
	\begin{array}{ccccc}\hline
\rowcolor{gray}	{\rm Mass \ Eigenstate} &\phantom{sp}& m^2 &\phantom{sp}& \Delta\\\hline
	 \delta U + \delta V && 36  && 8\\
\rowcolor{gray}	 \sqrt3\left( \,{\rm Im}\,\delta c^\Omega+{\rm Re}\,\delta b^\Omega\right) + 8 \, \delta U  && 36&& 8\\
\sqrt3\left({\rm Im} \,\delta c^\Omega -\, {\rm Re}\,\delta b^\Omega \right)+ \delta \phi && 27&&2(1+\sqrt{7})\\
\rowcolor{gray}	\sqrt3\left( {\rm Im}\,\delta b^\Omega +\, {\rm Re}\,\delta c^\Omega\right)-\delta C_0 && 27&&2(1+\sqrt{7})\\
\sqrt3\left({\rm Re}\,\delta b^\Omega + {\rm Im}\,\delta c^\Omega\right)-4\,\delta  U && 9&&2(1+\sqrt{3})\\
\rowcolor{gray}	\sqrt3\left(	{\rm Im}\,\delta c^\Omega - {\rm Re}\,\delta b^\Omega\right)- 2\,\delta \phi && 0 &&4\\
\sqrt3\, {\rm Re} \,\delta c^\Omega + \delta C_0&& 0&&4\\
\rowcolor{gray} \sqrt3\, {\rm Im}\,\delta b^\Omega +\delta C_0&& 0&&4\\\hline
	\end{array}
\end{equation}

We see that anomalous and irrational dimensions appear at this non-supersymmetric vacuum.
From the linear combinations of the scalar fields involved we also see that one should generically expect a mixing of the operators in the gauge theory. Clearly these operators cannot be written in terms of superfields, the vacuum being ${\cal N} = 0$.
Moreover, we notice that not only supersymmetry is completely broken at this vacuum, but also our gauge group, with all the 4 vector fields acquiring non-trivial masses
\begin{equation}
m^2 \;=\: \{\,36,\,18,\,18,\,9\,\}.
\end{equation}
It follows that their dual conformal operators also have irrational anomalous conformal dimension. Indeed, from the standard relation $\Delta = 2+\sqrt{1+ L^2m^2}$, we get
\begin{equation}
\Delta \;=\; \{\,2+\sqrt{33}\,, \,2+\sqrt{17}\,,\,2+\sqrt{17}\,, \,5\,\}.
\end{equation}

The masses (\ref{eq:NonSusyScalarMasses}) of the scalar fields in our truncation are all non-negative at this vacuum, and therefore obviously respect the Breitenlohner--Freedman (BF) bound required for stability.
However, in order to prove the full stability of this vacuum we should also compute the spectrum of fluctuations along directions orthogonal to our truncation, and this depends on the choice of the internal SE manifold.
For this reason, we cannot provide a general proof of (in)stability at this stage, but we are aware that in the case of the squashed $S^5$ manifold there are modes that develop an instability violating the BF bound \cite{Zaffa}.
Neglecting this issue for the time being, we will conclude our discussion by proposing some applications to the gauge/gravity correspondence, with a special emphasis on the (charged) domain-wall solutions interpolating between these vacua and their interpretation as Renormalization Group (RG) flows.

% subsection the_susy_breaking_vacuum (end)

\subsection{Flows from and between the vacua} % (fold)
\label{sub:flows_from_and_between_the_vacua}

It is interesting to notice that the two critical points of the potential can be parameterized by a single scalar field combination changing its expectation value. To discuss this, we define a consistent truncation of the 5-dimensional theory given in section \ref{sec:the_5d_action_and_scalar_potential} -- and therefore of type IIB supergravity -- which preserves the metric $g_{\mu\nu}$, the 1-form $A$, and sets (we chose $k=2$)
\begin{equation}\label{eq:FurtherRedAnsatz}
\begin{array}{rcl}
-U&=&\displaystyle V \;=\; {{\frac{1}{2}}}\log{\left(\cosh{\sigma}\right)},\\ [3mm]
c^\Omega &=& b^\Omega \,\tau = e^{\phi/2}\, e^{i \theta} \,\tau\, \tanh{\sigma},  \\ [2mm]
a_1^J &=& \displaystyle -A\,,
\end{array}
\end{equation}
where $\sigma,\,\theta$ parameterize the surviving complex scalar. 
The axio-dilaton $\tau=C_0+ie^{-\phi}$ is fixed to an arbitrary constant, while the remaining fields are set to zero.
We verified that this is a consistent truncation by plugging (\ref{eq:FurtherRedAnsatz}) into the 5-dimensional equations discussed in appendix \ref{sec:5d_equations_of_motion} as well as into the 5-dimensional action of section \ref{sec:the_5d_action_and_scalar_potential}, and by checking their compatibility. 
The truncated action reads
\begin{eqnarray}
S&=&\frac{1}{2\kappa_5^2}\int_M\Big[R*1-2d\sigma\wedge * d\sigma - \frac{1}{2}\sinh^2{(2\sigma)}\left(d\theta-3A\right)\wedge*\left(d\theta-3A\right) -\frac{3}{2}dA\wedge*dA  \nonumber\\[2mm]
&&\qquad\qquad+A\wedge dA\wedge dA \,-\, \mathcal V_{\rm eff} *1\Big{]}\,,
\label{truncatedaction}
\end{eqnarray}
where the truncated scalar potential is
\begin{equation}
	\mathcal V_{\rm eff} = 3 \cosh^2 \sigma \left[\cosh(2 \sigma) - 5\right].
\end{equation}
This consistent truncation has been previously presented in \cite{Gubser1} (see also \cite{KhavaevPilchWarner}).\footnote{Up to dilaton factors, the reduction ansatz of \cite{Gubser1} is recovered by identifying $\eta^{\rm there}=2\sigma^{\rm here}$.} Here we have shown how it can be embedded in our more general $\mathcal N=4$ reduction.

The critical points of $\mathcal V_{\rm eff}$ are at $\sigma = 0$, the supersymmetric one, and at $\sigma = \frac{1}{2}\log (2+\sqrt 3)$, the non-supersymmetric one.
Although the scalar potential contains only one scalar field, we stress that the kinetic term of the other scalar, $\theta$, vanishes at the supersymmetric critical point.
This means that one needs to perform a suitable field redefinition in order to obtain meaningful masses at that critical point.
After performing such a redefinition one can easily check that the resulting fluctuations have $m^2 = -3$.

\begin{figure}[ht]
	\centering
\includegraphics[scale=.35]{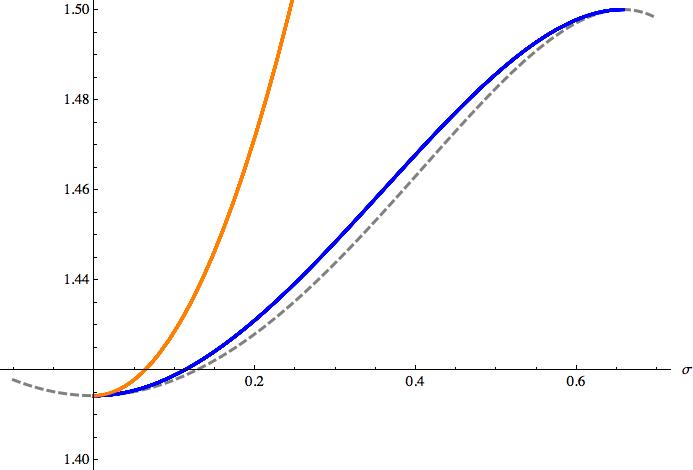} 
\caption{\small Plot of the numerical solutions for the superpotential (light orange) and fake superpotential (dark blue). 
The dashed gray line represents $\sqrt{|{\cal V}_{\rm eff}|/6}$ as a function of $\sigma$.}
\label{Figura}
\end{figure}

RG flows interpolating between the two dual conformal theories are domain-wall solutions of the 5-dimensional equations of motion supported by the scalar field $\sigma$ and possibly by some vector field if they are charged.
From the point of view of the dual field theory we expect such flows to arise when relevant deformations or vacuum expectation values of the operators are introduced.
For the uncharged case, the main relevant deformation involves the gaugino operator Tr $ W^2$.
Simply adding this operator to the dual conformal theory gives rise to the supersymmetric flow discussed in \cite{Girardello:1998pd} within 5-dimensional $\mathcal N=8$ supergravity, and lifted to type IIB supergravity in \cite{PilchWarner}.
This flow obviously overshoots the second critical point, which is non-supersymmetric, and ``flows to hades''.
On the other hand, there is always the option to turn on other operators and possibly also some vevs.
This corresponds to choosing specific initial conditions, which may lead to RG flows that may reach the second critical point and stop there.
The generic deformation will involve second-order differential equations \cite{Girardello:1998pd}. 
However, stable solutions (not necessarily BPS) will be constructed whenever the scalar potential can be written as \cite{Skenderis:1999mm}:
\begin{equation}
	\mathcal V_{\rm eff} \,=\, \frac94 (\partial_\sigma\mathcal W)^2 -6\mathcal  W^2 .
	\label{Veff}
\end{equation}
In this case, starting from a domain wall metric of the form
\begin{equation}
	ds^2 \,=\, e^{2 \rho(r)}ds^2(\mathbb R^{1,3}) + dr^2,
\end{equation}
it is straightforward to show that solutions to the first-order differential equations
\begin{equation}
	\rho' = \mathcal W\,, \qquad \sigma' = - \frac32\, \partial_\sigma \mathcal W\,, 
\end{equation}
are also solutions of the full equations of motion.
The supersymmetric solutions are described by the superpotential
\begin{equation}
	\mathcal W = \sqrt{2}\cosh^2 \sigma\,.
	\label{supo}
\end{equation}
However, this superpotential has only the supersymmetric critical point and hence cannot generate flows interpolating between the two vacua.
Since (\ref{Veff}) can be seen as a differential equation \emph{defining} the (fake) superpotential $\mathcal W$, we proceeded to its numerical integration, starting from both the critical points.
We display the result in figure~\ref{Figura}, where the  (light orange) numerical solution starting from the $\sigma = 0$, supersymmetric critical point, simply overlaps the analytic curve parameterized by (\ref{supo}).
The other numerical integration (dark blue) starts from the non-supersymmetric critical point and approaches the supersymmetric one for $\sigma \to 0$, within numerical error.
The corresponding first order flow for the scalar field gives the desired interpolating domain-wall.

A different option that can be considered is to allow for the domain-wall solution to be charged.
The corresponding dual solution describes the critical behaviour and the emergent relativistic conformal symmetry in superfluids or superconducting states of strongly coupled gauge theories \cite{Gubser2}.

% subsection flows_from_and_between_the_vacua (end)

\subsection{Further reductions} % (fold)

Besides the one presented in the above subsection, starting from the action in section \ref{sec:the_5d_action_and_scalar_potential} one can define further consistent truncations, encompassing several models previously studied in the literature.

As a first thing, one can consistently truncate the fields that are charged under the U(1) generated by $A$, namely set $b^\Omega \,=\, c^\Omega \,=\, a_1^\Omega \,=\, a_2^\Omega \,=\, 0\,$. This corresponds to expanding the higher-dimensional forms in terms of $J$ and $\eta$, excluding $\Omega$. Since $\eta$ and $J$ alone characterize a Sasakian structure, we believe that this reduction can be performed on any Sasaki manifold. It is then straightforward to reproduce the two non-supersymmetric consistent truncations derived in \cite{MaldMartTach}: the one with vector mass $m^2=8$ is obtained by setting to zero all the fields but $g_{\mu\nu},\,U,\,V,\,\phi\,,b^J,\,b_1$, while the one with vector mass $m^2=24$ arises from keeping just $g_{\mu\nu},\,U,\,V,\,A,\,a_1^J$.

Furthermore, by projecting out all the fields except $g_{\mu\nu}$ and $A$ (also setting $a_1^J =- A$), we get the consistent truncation to (the bosonic sector of) minimal $\mathcal N=2$ gauged supergravity studied in \cite{BuchelLiu}. 
It is also possible to truncate to $\mathcal N=2$ gauged supergravity coupled to matter. 
In particular, there exists a consistent truncation to $\mathcal N=2$ gauged supergravity with one hypermultiplet, corresponding to the intersection of our dimensional reduction and the $\mathcal N=8$ theory arising from the $S^5$ reduction. 
Indeed, if we take the identifications in (\ref{eq:FurtherRedAnsatz}) and in addition give dynamics to the axio-dilaton, we obtain the following extra pieces to the action (\ref{truncatedaction})
\begin{eqnarray}
S' &=&\frac{1}{2\kappa_5^2}\int_M\,\,\left[-\frac{1}{2}\,\cosh^2{\sigma}\,d\phi\wedge*d\phi-\frac{1}{2}e^{2\phi}\cosh^4{\sigma}\,dC_0\wedge*dC_0\right.\nonumber\\[2mm]
&&\quad\quad\quad\quad\left.+\frac{1}{2}e^\phi\sinh^2{(2\sigma)}\,dC_0\wedge*\left(d\theta-3A\right)\right].
\end{eqnarray}
These, together with the previous ones, complete the scalar $\sigma$-model to the $\nicefrac{\rm SU(2,1)}{\rm SU(2)\times U(1)}$ scalar manifold of the universal hypermultiplet.

\subsubsection*{Reducing to four dimensions}

Starting from the 5-dimensional theory obtained in section \ref{sec:the_5d_action_and_scalar_potential}, we can also go down to four dimensions by performing a circle reduction. 
In this way we provide a consistent truncation of type IIB supergravity on the particular SU(2) structure 6-dimensional manifold given by the direct product of a squashed Sasaki--Einstein manifold and $S^1$. 
The resulting \hbox{4-dimensional} theory is a gauged $\mathcal N=4$ supergravity with 3 vector multiplets and 20 scalars, some of which dualized to tensors. 
It is straightforward to determine the 4-dimensional gauging by applying to (\ref{eq:OurEmbTensor}) the map between the 4-dimensional and 5-dimensional embedding tensors provided in \cite{SchonWeidner}. 
The scalar potential is thus fixed, and it might be interesting to study some possible solutions of the theory. 
The ungauged supergravity obtained by switching off both the geometric and the RR fluxes corresponds to a consistent truncation of type IIB on $K3\times T^2$.

% subsection further_reductions (end)

% section discussion (end)

\bigskip

\section*{Acknowledgments}

\noindent We would like to thank D. Martelli, M. Petrini and especially A. Zaffaroni for stimulating discussions.
This work is supported in part by the ERC Advanced Grant no. 226455, \textit{``Supersymmetry, Quantum Gravity and Gauge Fields''} (\textit{SUPERFIELDS}), by the Fondazione Cariparo Excellence Grant {\em String-derived supergravities with branes and fluxes and their phenomenological implications} and by the European Programme UNILHC (contract PITN-GA-2009-237920).

%%%%%%%%%%%%%%%%%%%%%%%%%%%%%%%%%%%%%%%%%%%%%%%%%%%%%%%%%%%%%%
\appendix

%%%%%%%%%%%%%%%%%%%%%%%%%%%%%%%%%%%%%%%%%%%%%%%%%%%%%%%%%%%%%%
\section{Conventions} % (fold)
\label{sec:conventions}

The Hodge dual acting on the 10-dimensional vielbeine $E^A$ is defined as
\begin{equation}
*_{10}\, E^{A_1\ldots A_p} \,=\, {\textstyle{\frac{1}{(10-p)!}}}\,\epsilon^{A_1\ldots A_p}{}_{A_{p+1}\ldots A_{10}} \,E^{A_{p+1}\ldots A_{10}}\,,
\end{equation}
with $\epsilon_{01\ldots 9} = +1$. Analogous definitions hold for the lower-dimensional Hodge duals. Recall that in $d$ dimensions one has $** A_p = (-)^{p(d-p)+t}A_p$, where $t=0$ for euclidean signature and $t=1$ for lorentzian signature. 

Given a $p$-form $A_p$ and a $q$-form $B_q$ (with $p\leq q$), we define the $(q-p)$-form
\begin{equation}
A_p\lrcorner B_q := \frac{1}{p!(q-p)!} A^{M_1\ldots M_p}B_{M_1\ldots M_p M_{p+1}\ldots M_q}dx^{M_{p+1}}\wedge\cdots\wedge dx^{M_q} \;.
\end{equation}
Then we have the relation
\begin{equation}
A_p\wedge * B_q = (-)^{p(q-p)}*(A_p\lrcorner B_q) \,.
\end{equation}

Recalling (\ref{eq:*SEforms}), the reduction of the 10-dimensional Hodge dual to 5 dimensions leads to the following relations between a $p$--form $f_p$ on $M$ and its Hodge dual $*f_p$ on $M$:
\begin{equation}
\begin{array}{rcl}
	 *_{10}\, f_p & = & \frac{1}{2}e^{\frac{2p-2}{3}(4U+V)}\,(*f_p)\wedge J\wedge J\wedge(\eta + A),\\ [3mm]
	*_{10} \big[f_p\wedge(\eta + A) \big]& = & \frac{1}{2}(-)^{p+1}e^{\frac{2p-2}{3}(4U+V) - 2V}\,(*f_p)\wedge J\wedge J,\\[3mm]
	*_{10} \big[f_p\wedge J^{i} \big]& = & e^{\frac{2p-5}{3}(4U+V) + V}\,(*f_p)\wedge J^{i} \wedge(\eta + A),\\ [3mm]
	*_{10} \big[f_p\wedge J^{i}\wedge(\eta + A) \big]& = & (-)^{p+1} e^{\frac{2p-5}{3}(4U+V) - V}\,(*f_p)\wedge J^{i},\\ [3mm]
	*_{10} \big[f_p\wedge\frac12J^{i}\wedge J^{i} \big]& = & e^{\frac{2p-8}{3}(4U+V) + 2V}\,(*f_p) \wedge(\eta + A),\\ [3mm]
	*_{10} \big[f_p\wedge\frac12J^{i}\wedge J^{i} \wedge(\eta + A) \big]& = & (-)^{p+1} e^{\frac{2p-8}{3}(4U+V)}\,(*f_p)\,,
\end{array}
\label{eq:HodgeStars}
\end{equation}
where there is no sum on repeated indices.

% section conventions (end)

\section{Reduction of the equations of motion} % (fold)
\label{sec:5d_equations_of_motion}

In this appendix we briefly discuss the reduction of the 10-dimensional equations of motion and provide the full set of equations of motion for the 5-dimensional fields. It is crucial to remark that, thanks to the properties (\ref{eq:AlgConstr})--(\ref{eq:*SEforms}) of the expansion forms, once we plug our truncation ansatz into the 10-dimensional equations of motion the dependence on the internal coordinates drops out, so that the obtained equations are really 5-dimensional. 

The bosonic equations of motion of type IIB supergravity in the Einstein frame are 
\begin{eqnarray}
\nonumber R_{MN} &=& {\textstyle{\frac{1}{2}}}\,\partial_M \phi \partial_N \phi + {\textstyle{\frac{1}{2}}}\, e^{-\phi} (\iota_M H) \lrcorner (\iota_N H) + {\textstyle{\frac{1}{2}}}\,e^{2\phi}(F_1)_M(F_1)_N + {\textstyle{\frac{1}{2}}}\, e^{\phi} (\iota_M F_3) \lrcorner(\iota_N F_3)  \\ [2mm]
\label{eq:EoMgMN} && +\, {\textstyle{\frac{1}{4}}}\,(\iota_M F_5) \lrcorner(\iota_N F_5) - {\textstyle{\frac{1}{8}}}\, g_{MN}\left(e^{-\phi}H\lrcorner H + e^{\phi} F_3 \lrcorner F_3 \right) ,\\ [2mm]
\label{eq:EoMphi} d*d \phi &=& - {\textstyle{\frac{1}{2}}} e^{-\phi} H\wedge * H + e^{2\phi}F_1\wedge * F_1  + {\textstyle{\frac{1}{2}}} e^{\phi} F_3\wedge * F_3\,,\\ [2mm]
\label{eq:EoMB} d\big( e^{-\phi} *H \big) &=& e^\phi\, F_1\wedge * F_3 + F_3\wedge * F_5\,,\\ [2mm]
\label{eq:EoMC0} d\left(e^{2\phi} *F_1\right) &=& -e^{\phi}\, H\wedge * F_3, \\ [2mm]
\label{eq:EoMC2} d \left(e^{\phi} *F_3\right) &=& - H\wedge * F_5,\\ [2mm]
d*F_5 &=& H\wedge F_3\,.\label{IIBeom}
\end{eqnarray}
The equation of motion for $F_5$ and its Bianchi identity $dF_5=H\wedge F_3$ are actually equivalent, due to the self-duality constraint $ F_5\,=\, * F_5\,$.

The decomposition of the higher-dimensional Ricci tensor associated with the metric (\ref{eq:10dmetric}) was given in \cite{MaldMartTach} and we just reproduced their result. 
Translating the expressions (D.2)--(D.6) of \cite{MaldMartTach} to the 5-dimensional Einstein frame, we obtain (in flat indices)
\begin{eqnarray}\nonumber
	R^{(10)}_{ab} &=& e^{\frac{8}{3}U+\frac{2}{3}V}\left[R_{ab} -\frac{28}{3}\partial_a U\partial_b U -\frac{8}{3}\partial{}_{(a}U\partial{}_{b)}V -\frac{4}{3}\partial_a V \partial_b V \right.\\ [2mm]
	&& \left. \qquad \qquad-\;\frac{1}{2}e^{\frac{8}{3}U+\frac{8}{3}V}F_{ac} F_b{}^c  \,+\, \frac{1}{3}\eta_{ab}\, \square_5 (4U+V) \right],\label{eq:Ricci_ab}\\ [2mm]
	R^{(10)}_{ij} &=& \delta_{ij} \left[ 6\,e^{-2U} - 2\,e^{-4U+2V} - e^{\frac{8}{3}U+\frac{2}{3}V}\square_5 U \right],\label{eq:Ricci_ij}\\ [2mm]
	R_{99}^{(10)} &=& 4\,e^{-4U+2V} - e^{\frac{8}{3}U+\frac{2}{3}V} \square_5 V +\frac{1}{4}e^{\frac{16}{3}U+\frac{10}{3}V}F_{ab}F^{ab},\\ [2mm]
	R^{(10)}_{ai} &=& R^{(10)}_{i9} \;\;=\;\; 0,\\ [2mm]
	R^{(10)}_{a9} &=& -\frac{1}{2} e^{\frac{4}{3}U - \frac{2}{3}V} \nabla^b\left(e^{\frac{8}{3}U + \frac{8}{3}V}F_{ba}\right)\label{eq:Ricci_a9} ,
\end{eqnarray}
where we denote $F=dA$.
The Ricci tensor on the left hand side of the above equations is expressed in terms of the 10-dimensional vielbeins $E_A$, namely $R^{(10)}_{AB}= R^{(10)}_{MN}E^M_AE^N_B$, whereas the vielbeins employed on the right hand side are the ones of the 5-dimensional Einstein metric $g_{\mu\nu}$, defining $ds^2(M)$ in (\ref{eq:10dmetric}). Here, the $a,b$ indices are flat indices on $M$, while $i,j$ are flat indices on the K\"ahler--Einstein base $B_{\rm KE}$ of our internal manifold.
	
Recalling the expressions in section \ref{sec:iib_supergravity_reduced_on_se__5} for the reduction of the various 10-dimensional form fields, the reduction of the dilaton equation (\ref{eq:EoMphi}) is straightforward, and yields 
\begin{eqnarray}
\nonumber d*d\phi \!\!&-&\!\! e^{2\phi}dC_0\wedge * dC_0+\frac{1}{2}e^{-\phi}\left[ e^{\frac{16}{3}U+ \frac{4}{3}V}h_3\wedge *h_3 + e^{\frac{8}{3}U-\frac{4}{3}V}h_2\wedge*h_2 + 2e^{-4U}h_1^J\wedge *h_1^J\right. \\ [2mm] 
&+&\!\! \left. 2e^{-4U}{\rm Re}\left(  h_1^\Omega \wedge * \overline{h_1^\Omega}\right)  \,+\, 2e^{-\frac{20}{3}U-\frac{8}{3}V}| h_0^\Omega|^2 *1  \right]\;-\;\frac{1}{2}e^{\phi}\Big[h_p\to g_p\Big]\;=\;0\,,
\end{eqnarray}
which is consistent with the 5-dimensional action of section \ref{sec:the_5d_action_and_scalar_potential}. By $h_p\to g_p$ we denote repetition of the terms in the previous parenthesis with $h_p$ replaced by $g_p$.

Equation (\ref{eq:EoMC0}) yields the 5-dimensional equation of motion for the RR axion $C_0(x)$:
\begin{eqnarray}
\nonumber d\left(e^{2\phi}*dC_0\right) \!\!&+&\!\! e^{\phi}\left[ e^{\frac{16}{3}U+ \frac{4}{3}V}h_3\wedge *g_3 + e^{\frac{8}{3}U-\frac{4}{3}V}h_2\wedge*g_2 + 2e^{-4U}h_1^J\wedge *g_1^J\right. \\ [2mm] 
&& \quad+ \left. 2\,{\rm Re}\left( e^{-4U} h_1^\Omega \wedge * \overline{g_1^\Omega} \,+\, e^{-\frac{20}{3}U-\frac{8}{3}V} h_0^\Omega \overline{g_0^\Omega} *1\right)   \right]\;=\;0\,.
\end{eqnarray}

The $H$-equation of motion (\ref{eq:EoMB}) gives the following four 5-dimensional expressions
\begin{eqnarray}\nonumber
d\left( e^{\frac{16}{3}U + \frac{4}{3}V -\phi}*h_3 \right ) &=& e^{\frac{16}{3}U + \frac{4}{3}V+\phi}dC_0\wedge *g_3 + 2 g_3f_0 - 2 g_2\wedge f_1 + 2g_1^J\wedge f_2^J \\ [2mm] 
&& + \,2\,{\rm Re} \left(g_1^\Omega \wedge\overline{f_2^\Omega} - g_0^\Omega\, \overline{f_3^\Omega} \right),\label{eq:b2eq}
\end{eqnarray}
which is the equation of motion for $b_2\,$, 
\begin{eqnarray}
&&\!\!\!\!\! d\left(e^{\frac{8}{3}U-\frac{4}{3}V-\phi}*h_2 \right)- 4e^{-4U-\phi} *h_1^J - e^{\frac{16}{3}U+\frac{4}{3}V-\phi}dA\wedge *h_3 \,=\, \\ [2mm]
\nonumber\!\!&=&\!\! e^{\frac{8}{3}U-\frac{4}{3}V+\phi}dC_0\wedge *g_2 - 2g_3\wedge f_1 \,+\,2 e^{-\frac{4}{3}U-\frac{4}{3}V}g_1^J\wedge *f_2^J  +\,2\,e^{-\frac{4}{3}U-\frac{4}{3}V}{\rm Re}\left( g_1^\Omega\wedge * \overline{f_2^\Omega} \right),
\end{eqnarray}
which is the equation of motion for $b_1\,$, 
\begin{equation}
d\left( e^{-4U-\phi} *h_1^J \right) \;=\; e^{-4U+\phi}dC_0\wedge *g_1^J +  g_3\wedge f_2^J + e^{-\frac{4}{3}U-\frac{4}{3}V} g_2\wedge *f_2^J + 2e^{-8U} g_1^J \wedge * f_1,
\end{equation}
which is the equation of motion for $b^J$, and
\begin{eqnarray}\nonumber
D(e^{-4U-\phi}*h_1^\Omega) \!\!\!&=&\!\!\! e^{-4U+\phi}dC_0\wedge *g_1^\Omega + g_3\wedge f_2^\Omega + e^{-\frac{4}{3}U-\frac{4}{3}V} g_2\wedge *f_2^\Omega + 2e^{-8U} g_1^\Omega \wedge * f_1 -\\ [3mm]
&&\qquad\qquad\qquad - \left( 3i\,e^{-\frac{20}{3}U-\frac{8}{3}V-\phi}h_0^\Omega - 2 e^{-\frac{32}{3}U-\frac{8}{3}V}g_0^\Omega f_0\right) *1 \,,\label{eq:bOmEq}
\end{eqnarray}
which is the equation of motion for $\overline{b^\Omega}$.
Once more, all these equations are compatible with the ones obtained from the action of section \ref{sec:the_5d_action_and_scalar_potential}.

We now consider the $C_2$ equation given in (\ref{eq:EoMC2}). 
This is obtained from (\ref{eq:EoMB}) by performing $H\to F_3$, $ F_3\to -H$, $F_1\to 0$ and $-\phi \,\to\, \phi$.
It follows that the corresponding 5-dimensional equations, to be interpreted as equations of motion for $c_2$, $c_1$, $c^J$, $\overline{c^\Omega}\,$, are derived from the equations above respectively for $b_2$, $b_1$, $b^J$, $\overline{b^\Omega}$ by implementing
\begin{equation}\label{eq:Transforms}
h_p\to g_p\quad,\quad g_p \to -h_p \quad,\quad dC_0\to 0 \quad,\quad -\phi \,\to\, \phi \;.
\end{equation}

The equations of motion coming from the reduction of the 5-form are also its Bianchi identities, following the discussion in section \ref{sub:the_self_dual_5_form}.
The equation of motion of $a_1^J$ is
\begin{equation}
d\left(e^{-\frac{4}{3}U-\frac{4}{3}V}*f_2^J\right)-4e^{-8U}*f_1-f_2^J\wedge dA\,=\,Dc^J\wedge\left(db_2-b_1\wedge dA\right)- \,\,b\leftrightarrow c\,.
\label{f2Jeom}
\end{equation}
As $f_1$ defines the covariant curvature for the field $a$, which is a pure gauge of the gauge symmetry inherited from shifting the $C_4$ potential, its equation of motion is not giving us new independent information on the dynamics.
In fact it can be obtained by exterior differentiation on (\ref{f2Jeom}):
\begin{equation}
	d\left(2e^{-8U}*f_1\right) \,=\, \left(db_2-b_1\wedge dA\right)\wedge dc_1-\,\,b\leftrightarrow c\,.
\end{equation}
Finally, the equation of motion of $a_2^\Omega$ reads
\begin{equation}
D a_2^\Omega - a_1^\Omega\wedge dA + \frac12\left[b_2 \wedge Dc^\Omega + b^\Omega (dc_2-c_1\wedge dA) -\,\,b\leftrightarrow c\right] = -e^{-\frac{4}{3}U - \frac{4}{3}V}* f_2^\Omega,
\end{equation}
and is equivalent to the duality relation between $f_2^\Omega$ and $f_3^\Omega$ given in (\ref{eq:Dualityf_p}). The equation for $a_1^\Omega$ is just its covariant derivative, reflecting the fact that $a_1^\Omega$ is a pure gauge degree of freedom.

We are now left with the reduction of the higher-dimensional Einstein equation (\ref{eq:EoMgMN}). 
Also recalling eq. (\ref{eq:Ricci_ij}), we see that the block with $ij$ indices is proportional to $\delta_{ij}$ and therefore yields a single scalar equation, which reads
\begin{eqnarray}\nonumber
&& \square U -6 e^{-\frac{14}{3}U-\frac{2}{3}V} +2e^{-\frac{20}{3}U+\frac{4}{3}V} + \frac{e^{-\phi}}{4}\left[ -\frac{1}{2}e^{\frac{16}{3}U+\frac{4}{3}V}h_3\lrcorner h_3 - \frac{1}{2}e^{\frac{8}{3}U-\frac{4}{3}V}h_2\lrcorner h_2 + e^{-4U}h_1^J\lrcorner h_1^J  \right. \\ [2mm]
&& + e^{-4U}h_1^\Omega\lrcorner\overline{h_1^\Omega} +e^{-\frac{20}{3}U-\frac{8}{3}V}|h_0^\Omega|^2 \bigg] +\frac{e^{\phi}}{4}\bigg[ h_p\to g_p\bigg]+e^{-8U}f_1\lrcorner f_1 + e^{-\frac{32}{3}U-\frac{8}{3}V}f_0^2 \,=\, 0,
\label{eq:Einstij}
\end{eqnarray}
where the metric involved in the D'Alembertian and in the contraction of the indices is $g_{\mu\nu}$.
The $9\,9$ component of the same 10-dimensional equation reads
\begin{eqnarray}\nonumber
&&\square V -4 e^{-\frac{20}{3}U+\frac{4}{3}V} - \frac{1}{2} e^{\frac{8}{3}U+\frac{8}{3}V}F\lrcorner F + \frac{e^{-\phi}}{4}\left[ -\frac{1}{2}e^{\frac{16}{3}U+\frac{4}{3}V}h_3\lrcorner h_3 +\frac{3}{2}e^{\frac{8}{3}U-\frac{4}{3}V}h_2\lrcorner h_2 \right.\\ [2mm]
\nonumber &&  - e^{-4U}h_1^J\lrcorner h_1^J -e^{-4U}h_1^\Omega\lrcorner\overline{h_1^\Omega} +3e^{-\frac{20}{3}U-\frac{8}{3}V}|h_0^\Omega|^2 \bigg] + \frac{e^{\phi}}{4}\bigg[ h_p\to g_p\bigg]\\ [2mm]
\label{eq:Einst99}&& -e^{-8U}f_1\lrcorner f_1 +\frac{1}{2}e^{-\frac{4}{3}U-\frac{4}{3}V}f_2^J\lrcorner f_2^J + \frac{1}{2} e^{-\frac{4}{3}U-\frac{4}{3}V}f_2^\Omega\lrcorner \overline{f_2^\Omega} + e^{-\frac{32}{3}U-\frac{8}{3}V}f_0^2 \;\;=\;\; 0\,.
\end{eqnarray}
The two 5-dimensional scalar equations above are equivalent to the equations of motion for $U$ and $V$.

We get no 5-dimensional equations from the Einstein equations with $ai$ or with $i \,9$ indices, because all the terms appearing there separately vanish within our reduction ansatz.

The Einstein equation with $a\,9$ flat indices reduces to
\begin{eqnarray}\nonumber
&&-d\left(e^{\frac{8}{3}U+\frac{8}{3}V}*dA\right) + e^{-\phi}\left[e^{\frac{16}{3}U+\frac{4}{3}V}h_2\wedge *h_3 + 2e^{-4U}{\rm Re}\left(h_0^\Omega *\overline{h_1^\Omega}\right)  \right]   +e^{\phi}\Big[ h_p\to g_p\Big]\\ [2mm]
&& + f_2^J\wedge f_2^J + f_2^\Omega\wedge \overline{f_2^\Omega} + 4\,e^{-8U}f_0*f_1 \;=\;0\,,
\end{eqnarray}
which is the equation of motion for $A$ and is equivalent to the one derived from the action presented in section \ref{sec:the_5d_action_and_scalar_potential}.

Finally we study the $ab$ components of the Einstein equation, where we employ (\ref{eq:Einstij}), (\ref{eq:Einst99}) to get rid of the $\square (4U+V)$ term appearing in the expression (\ref{eq:Ricci_ab}) for the higher-dimensional Ricci tensor. 
This yields the 5-dimensional Einstein equation
\begin{eqnarray}\nonumber
R_{ab} \!&=&\!\!\frac{28}{3}\partial_a U\partial_b U +\frac{8}{3}\partial_{(a}U\partial_{b)}V +\frac{4}{3}\partial_a V\partial_b V +\frac{1}{2}e^{\frac{8}{3}U + \frac{8}{3}V}\iota_aF\lrcorner\, \iota_bF + \frac{1}{2}\partial_a\phi\partial_b \phi +\frac{e^{2\phi}}{2}\partial_aC_0\partial_bC_0\\ [2mm]
\nonumber \!\!\!&+&\!\!\!\frac{1}{2}e^{-\phi}\left[ e^{\frac{16}{3}U + \frac{4}{3}V}\iota_a h_3\lrcorner \,\iota_b h_3  + e^{\frac{8}{3}U - \frac{4}{3}V}\iota_a h_2\lrcorner \,\iota_b h_2 + 2e^{-4U} h_{1\,a}^J h_{1\,b}^J + 2e^{-4U} h_{1(a}^\Omega \overline{h{}_{1\,b)}^\Omega}\,\right]\\ [2mm]
\nonumber \!\!\!&+&\!\!\!\frac{1}{2}e^{\phi}\Big[ h_p \to g_p \Big] \, +\, e^{-\frac{4}{3}U-  \frac{4}{3}V} \iota_a f_2^J\lrcorner\, \iota_b f_2^J + e^{-\frac{4}{3}U-  \frac{4}{3}V} \iota_{(a} f_2^\Omega\lrcorner \,\iota_{b)} \overline{f_2^\Omega} + 2e^{-8U}f_{1\,a} f_{1\,b}\\ [2mm]
\nonumber  \!\!\!&-&\!\!\! \frac{1}{3}\eta_{ab}\left[ 24\,e^{-\frac{14}{3}U-\frac{2}{3}V} -4\,e^{-\frac{20}{3}U+\frac{4}{3}V} +\frac{1}{2}e^{\frac{8}{3}U+\frac{8}{3}V}F\lrcorner F \right.\\ [2mm]
\nonumber \!\!\!&+&\!\!\!\left. e^{-\phi}\left( e^{\frac{16}{3}U+\frac{4}{3}V}h_3\lrcorner h_3  + \frac{1}{2}e^{\frac{8}{3}U-\frac{4}{3}V}h_2\lrcorner h_2 - e^{-\frac{20}{3}U-\frac{8}{3}V}|h_0^\Omega|^2 \right) + e^{\phi}\bigg(h_p\to g_p\bigg)\right.\\ [2mm]
 \!\!\!&+&\!\!\! \left. e^{-\frac{4}{3}U-\frac{4}{3}V}\left(f_2^J \lrcorner f_2^J + f_2^\Omega \lrcorner \overline{f_2^\Omega} \right) - 2\,e^{-\frac{32}{3}U-\frac{8}{3}V}f_0^2  \!\!\phantom{\frac{1}{1}}\right],
\end{eqnarray}
which also matches the one obtained by varying the action of section \ref{sec:the_5d_action_and_scalar_potential}.

% section 5d_equations_of_motion (end)

%%%%%%%%%%%%%%%%%%%%%%%%%%%%%%%%%%%%%%%%%%%%%%%%%%%%%%%%%%%%%%

\end{document}